\documentclass[aps,prx,reprint,showpacs,superscriptaddress,floatfix]{revtex4-2}
\usepackage{comment}
\usepackage{graphics,amssymb,amsmath,epsfig,color,textgreek}
\usepackage{graphicx}
\usepackage[dvipsnames]{xcolor}
\usepackage{dcolumn}
\usepackage{bm}
\usepackage[colorlinks=true,citecolor=cyan]{hyperref}
\hypersetup{colorlinks=true,citecolor=cyan,linkcolor=red,urlcolor=magenta}
\usepackage{braket}
\usepackage[normalem]{ulem}
\usepackage{cancel}
\usepackage{diagbox}
\usepackage{lipsum}
\usepackage{algorithm}
\usepackage[noend]{algpseudocode}
\usepackage{cleveref}
\usepackage{orcidlink}
\usepackage[mathscr]{eucal}
\usepackage{booktabs}
\usepackage{placeins}

\DeclareMathOperator*{\argmin}{arg\,min}

\crefformat{figure}{Fig.~#2#1#3}
\crefformat{equation}{Eq.~#2#1#3}
\crefformat{appendix}{App.~#2#1#3}
\crefformat{section}{Sec.~#2#1#3}

\newcommand{\mbf}[1]{\mathbf{#1}}
\newcommand{\mbfI}{\mathbf{I}}
\newcommand{\mbfA}{\mathbf{A}}
\newcommand{\mbfB}{\mathbf{B}}
\renewcommand{\c}{\hat{c}^\phant}
\newcommand{\cd}{\hat{c}^\dagger}
\newcommand{\mbfH}{\mathbf{H}}
\newcommand{\mbfV}{\mathbf{V}}
\newcommand{\mbfW}{\mathbf{W}}
\newcommand{\reals}{\mathbb{R}}
\newcommand{\complex}{\mathbb{C}}

\newcommand{\mcH}{\boldsymbol{\mathcal{H}}}
\newcommand{\mcB}{\boldsymbol{\mathcal{B}}}
\newcommand{\param}{\vec{p}}
\newcommand{\relerror}{\mathcal{E}^\text{rel}}
\newcommand{\phant}{{\phantom{\dagger}}}

\newcommand{\Dh}{\Delta\boldsymbol{\mathcal{H}}}
\newcommand{\off}{\mathrm{off}}
\newcommand{\on}{\mathrm{on}}

\begin{document}

\title{Compact representation of strongly correlated Green's functions:\\the MOR+EC way to explore phase space}

\author{Norman~Hogan\orcidlink{0000-0003-0720-4949}}
\email{anhogan3@ncsu.edu}
\affiliation{Department of Physics and Astronomy, North Carolina State University, Raleigh, North Carolina 27695, USA}

\author{A.~F.~Kemper\orcidlink{0000-0002-5426-5181}}
\email{akemper@ncsu.edu}
\affiliation{Department of Physics and Astronomy, North Carolina State University, Raleigh, North Carolina 27695, USA}

\author{Carlos~Mejuto-Zaera\orcidlink{0000-0001-5921-0959}}
\email{cmejutozaera@irsamc.ups-tlse.fr}
\affiliation{Univ Toulouse, CNRS, Laboratoire de Physique Théorique, Toulouse, France}

\date{October 21, 2026}

\begin{abstract}
Repeated evaluation of the single-particle Green's function (GF) across parameter space is a recurring bottleneck in many-body methods, limiting the resolution at which phase boundaries may be probed and the frequency resolution of computed spectra. We introduce MOR+EC, a framework that constructs reusable reduced-order models for computing single-particle Green's functions by combining eigenvector continuation (EC) for parameter space exploration and model order reduction (MOR) for extrapolation in frequency space. Contrary to conventional parameterized MOR, we construct these reduced-order models with a parameter-independent resolvent, further reducing the number of required full-space evaluations. Benchmarking against exact diagonalization as the impurity solver for our example case of DMFT calculations for single- and two-band models, MOR+EC reproduces the DMFT-converged impurity GF to an average relative error below $10^{-3}$, with a median wall-time speedup of $9.5$--$92\times$. This accuracy and efficiency together resolve a fine-grained scan of the impurity occupation as doping is tuned and produce a high-resolution phase diagram of an orbital-selective Mott transition. The reduced-order model also reproduces real- and imaginary-frequency spectra at no additional cost in full-space evaluations. This framework applies broadly to GF-based methods requiring repeated parametric evaluation and high resolution of the frequency axis.
\end{abstract}

\maketitle

\section{Introduction}

\begin{figure*}
    \centering
    \includegraphics[width=0.85\linewidth]{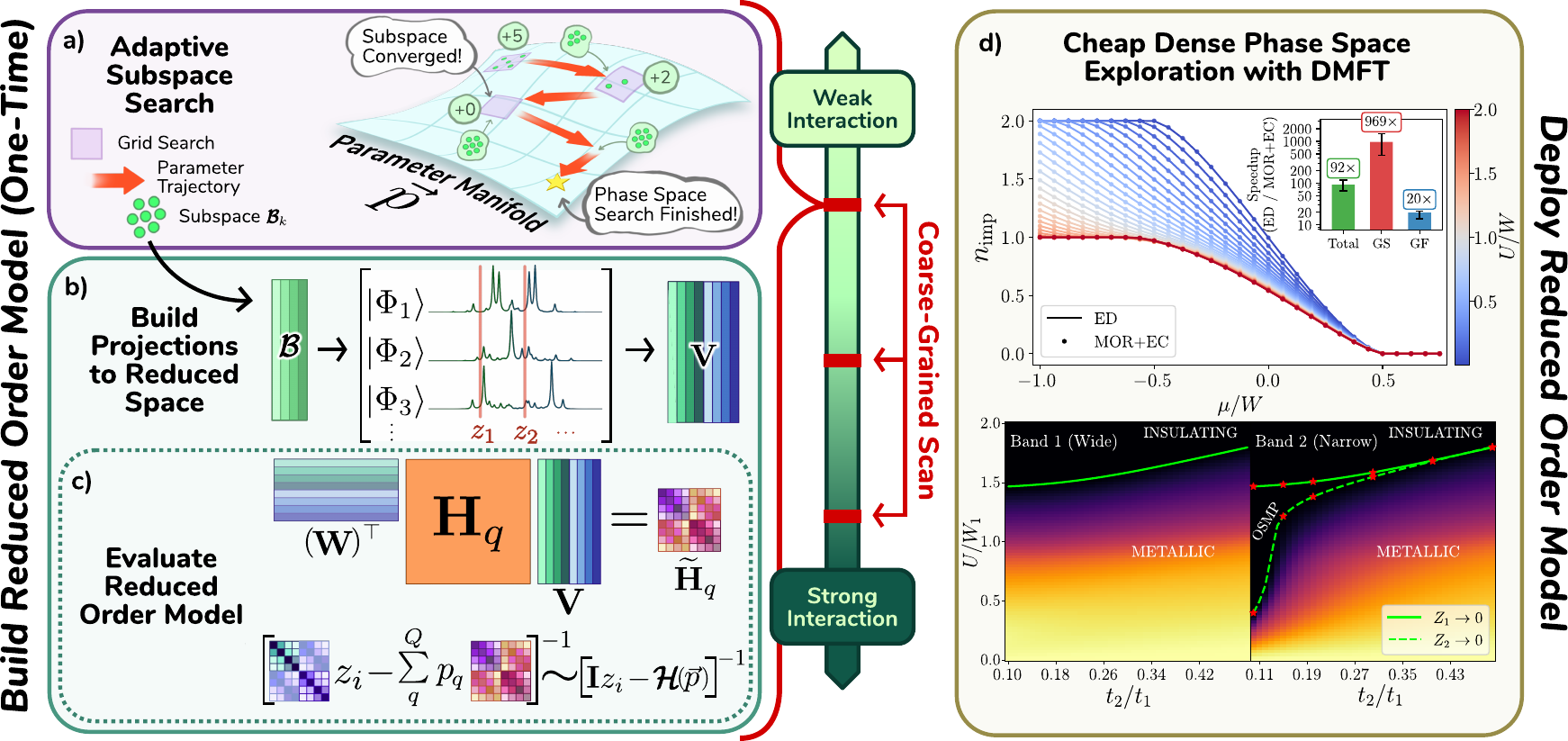}
    \caption{\textbf{Summary of this work.} We combine a reduced basis method for ground state determination [eigenvector continuation (EC)] and Green's function (GF) evaluation [model order reduction (MOR)] to build (a-c) and deploy (d) reduced-order models for dense phase space exploration, and apply these methods to dynamical mean field theory (DMFT). (a) Using EC, we build a reduced basis $\mcB$ to represent the ground state across the parameter manifold using global and local grid searches during DMFT calculations. (b) Using the subspace $\mcB$ as interpolation directions, we choose some interpolation frequencies $z_i$ that build a compact representation of the impurity GF using projections ($\mbfV,\mbfW$) built from a collection of these response snapshots. (c) These projections take all full-order operators to our reduced-order model, allowing for a two-fold speedup: ground state determination in the EC subspace $\mcB$ and GF evaluation with MOR. (d) After the initial investment in building these reduced-order models, dense phase space exploration is made significantly more efficient, allowing for a fine-grained scan of phase diagrams with a significant speedup in the solver evaluation time.}
    \label{fig: summary}
    
\vspace{-6pt}
\end{figure*}

Phase space exploration is a central task in solid state physics and materials science, and an indispensable ingredient for theory-assisted design of functional materials with targeted opto-electronic properties.
It requires solving the Schr\"odinger equation accurately and efficiently over many atomic configurations or chemical compositions, yielding not only a reasonable approximation for the ground state energy, but often also information on excited states or response functions. For those properties governed by single-particle excitations, the one-body Green's function (GF) is an appropriate target~\cite{martin_reining_ceperley_2016}; only when it is accessible in a streamlined manner can a truly high-throughput material exploration be performed.

While nowadays such ambitious computational screenings are possible through Kohn-Sham density functional theory DFT~\cite{Jones2015}, current functional approximations preclude its application to strongly correlated materials~\cite{Martin2008}. These have electronic dynamics that are dominated by multiple, competing energy scales~\cite{QuantumMatEditorial}, typically between the kinetic energy and Coulomb repulsion, which tend to electron delocalization/localization, respectively. That competition is what makes them attractive for functional material design: a single strongly correlated material can host a wide palette of exotic phases of matter, tuning between them through slight changes in external parameters like temperature, pressure, or doping~\cite{Lee2006}. Capturing this accurately requires methods beyond DFT, and the price to pay is an increased computational cost -- the many-body Hilbert space, and with it the algorithmic complexity, scales exponentially with the system size regardless of the approach used.  This prevents any kind of high-throughput calculations, as each point in configuration space or in the phase diagram requires a fresh full evaluation of the quantities of interest: ground state properties, or responses to external fields.

An increasingly popular and successful strategy is given by the quantum embedding framework~\cite{Kotliar1996,Kotliar2006,Sun2016}, which maps the in principle intractable strongly correlated material of interest into a smaller, computationally amenable impurity model. This model represents a small subset of the degrees of freedom of the original system, coupled to a non-interacting bath, and the mapping is determined self-consistently such that the impurity model faithfully reproduces some observable of interest.
When the observable is the GF, one typically resorts to the dynamical mean-field theory (DMFT)~\cite{Kotliar1996} or self-energy embedding theory~\cite{Zgid2015,Lan2015,Nguyen2016,Lan2017,Rusakov2018,Iskakov2020}; DMFT in particular has been extensively used to describe correlated solids through models and \emph{ab initio} descriptions alike~\cite{Kotliar2006,Paul2019}. Despite its great success, DMFT is still computationally demanding: it requires the repeated evaluation of impurity model GFs in order to find the bath that most faithfully represents the system of interest. The phase space exploration problem is thus a nested one: it enters the design of tailored materials, but also the solution of a single material configuration within its auxiliary, embedded impurity model description.

For the inner problem, the ideal impurity solver would be black-box, such that computational explorations can be automated without the need for human monitoring. Recent years have seen great advancements in this regard, e.g. machine learning-based methods~\cite{Sheridan2021,rigo2025,Zhouyin2026,Giuli2026} or the connector approach~\cite{Aouina2020,Vanzini2022}. Among these, the Eigenvector Continuation (EC) literature deserves special mention due to its simplicity and efficiency~\cite{frame2018EC,francis2022subspacediagonalizationquantumcomputers,herbst2022surrogatemodels,Mejuto-Zaera2023quantumEC,duget2024EC,Atalar2024,agrawal2025EC,Rath2025,hoganEfficientQuantumImplementation2026}: essentially, it proposes extremely low-rank representations of parametrized Hamiltonians through a variationally optimal ground state expansion.
Nevertheless, most of these techniques focus on reproducing static expectation values, and typically cannot access the GF accurately in the strongly correlated limit.

One relatively underexplored strategy for reaching the GF itself is model order reduction (MOR)~\cite{antoulas2020interpolatory,Opmeer2012MORrationalinterp,aretzNestedOperatorInference2025,peherstorferDatadrivenOperatorInference2016}, widely used in engineering and recently applied to problems such as absorption spectra~\cite{roel2017MOR}. MOR constructs a reduced-order model from a limited set of training inputs -- requiring the full system to be solved for only those -- enabling rapid approximation of the full model's response, here the GF. Extensions such as parameterized MOR (PMOR)~\cite{baur2011pmor,benner2016pmorsurvey,balickiMultivariateRationalApproximation2025,ionitaDataDrivenParametrizedModel2014} allow generalization across parameter regimes. The cost at this construction stage is amortized over subsequent evaluations: with sufficiently representative training data, reduced-order models reproduce full-order results with high fidelity at much lower cost.

In this work, we demonstrate the effectiveness of this strategy, combined with EC, for evaluating GF of correlated models accurately and efficiently. We construct a low-rank representation for the ground state subspace using
EC~\cite{duget2024EC}, forming an interpolation manifold in parameter space (see~\cref{fig: summary}(a,b)),
and, building on this, a reduced-order model (ROM) for the impurity GF by
interpolating across spectral frequencies using a minimal set of inputs (see~\cref{fig: summary}(c)). The problem is thus separated into two complementary reductions: one in parameter space, via the ground state subspace, and one in frequency space, via the GF construction. Applied to DMFT, this yields high-fidelity ROMs for single- and two-band impurity models which produce ED quality results at median wall-time speedups of $92\times$ and $9.5\times$ respectively (see~\cref{fig: summary}(d)). We thereby address the nested nature of the phase space exploration problem in strongly correlated materials, and the resulting models are made available as reusable tools for future studies.

The idea of constructing \textit{once-and-for-all} ROMs for impurity solvers has recently been explored in~\cite{Giuli2026} within the ghost Gutzwiller embedding~\cite{Lanata2017,Mejuto2023a,Mejuto2024,Giuli2025,Tagliente2025,Pasqua2026,Giuli2026b,Tagliente2026,Pasqua2026b,Giuli2026c}, where the focus is on the one-body reduced density matrix of the impurity model. By developing a ROM capable of extracting the frequency-dependent GF, we extend the applicability of this family of methods to DMFT and, more broadly, to any problem requiring high-resolution GFs
across parameter regimes -- the GF-based embedding methods named above among them. In such settings, investing in high-fidelity, reusable ROMs may be more efficient than repeatedly solving full-order models, enabling faster benchmarking, phase space exploration, and large-scale simulations.

The structure of this paper is as follows: in~\cref{sec: morec in brief}, we provide a broad description of our approach of combining MOR with EC, which we have named MOR+EC. In~\cref{sec: methods}, we first provide a definition of the problem statement (Sec.~\ref{sec: setup and notation}), then follow this with details on EC and MOR on their own (Sec.~\ref{sec: eigenvector continuation} and Sec.~\ref{sec: ROM by interp}). Then we explain how we combine these methods (Sec.~\ref{sec: MOREC}) for the DMFT application (Sec.~\ref{sec: DMFT}). We then showcase our results in~\cref{sec: Results} and discuss the implications of MOR+EC, its limitations, and future directions in~\cref{sec: discussion}.

\section{MOR+EC in brief}\label{sec: morec in brief}

We summarize here the complete MOR+EC construction, so that it may be read independently of the derivations that follow. \cref{sec: methods} develops each step, and the appendices collect the implementation details.

MOR+EC requires three ingredients: a Hamiltonian $\mcH(\param)=\sum_q^Q p_q \mbfH_q$ where the parameters $\param$ enter only as scaled coefficients for a fixed set of operators $\mbfH_q$; a solver capable of returning a ground state $|\Psi(\param)\rangle$ at a perscribed $\param$ and of applying the resolvent $(\mbfI z-\mcH)^{-1}$ to a given state; and a parameter domain to be explored. The output is a collection of small matrices: $\mbf{h}_q\in \reals^{k\times k}$ and $\widetilde\mbfH_q\in \complex^{m\times m}$, one pair per Hamiltonian term. These are the ROMs for the ground state and the Green's function, respectively. From each, the Hamiltonian's ground state and single-particle GF are obtained at any ($\param$,$z$) within that domain at a cost independent of the full-space many-body dimension $d\gg m > k$.

The crux is thus building the small matrices $\mbf{h}_q\in \reals^{k\times k}$ and $\widetilde\mbfH_q\in \complex^{m\times m}$.
This is done by sampling parameter and frequency space ($\param$,$z$) to choose a judicious set of eigenstates $|\Psi(\param)\rangle$ onto which to project the full Hamiltonian operator.
The $\mbf{h}_q\in \reals^{k\times k}$ matrices are direct projections of the $\mbfH_q$ operators onto these eigenstates, and result in a compact yet faithful representation from which to extract reliable eigenenergies and static expectation values.
The $\widetilde\mbfH_q\in \complex^{m\times m}$ matrices include additional information of the resolvents $(\mbfI z-\mcH)^{-1}$ in a selected frequency grid, and thus can be used as compact representations for computing Green's functions, much like the Krylov Hamiltonian in the Lanczos algorithm

The procedure separates into two stages: construction of the ROMs, performed once, and evaluating the ground state and GF using this ROM for every parameter point in a specified domain. The construction stage is dominated by at least $k$ full-space ground state solves and the $r$ factorizations of resolvents at selected frequencies $\mbf{z}\in\{z_i\}_{i=1}^r$, while the evaluation over the domain requires no reference to the full-space dimension at all. Two features distinguish our approach from standard parametric MOR. First, the EC basis $\mcB$ supplies the $k$ interpolation directions rather than a separate greedy search over $\param$, allowing the parameter and frequency spaces to share a single set of ground state solves. Second, all resolvents are evaluated at a fixed Hamiltonian $\mcH_0$, so that only $r$ shifted operators must be factored instead of $kr$ with each factorization being reused across the $k$ columns of $\mcB$. The consequences of this second choice are examined in Appendix~\ref{asec: param indep resolvent}. A final compression step (Appendix~\ref{asec: compression}) projects the ROM onto the directions its solutions occupy along the scan, lowering $m$ without further full-space operations.

For the two-band impurity model considered in our application of the methods, these steps reduce the ground state sector of dimension $d_\mathrm{GS}=4900$ to a basis of order $k=108$. For evaluations of the single-particle GF for this model, which involve particle/hole sectors of dimension $d_{p/h}=3920$, we find a ROM of $m=305$, compressed to $m'\approx90$, lowering the cost of a single call to the impurity solver from $\sim 63$\,ms to $\sim5$\,ms. For the single-band model, the offline investment is recovered after approximately 1040 solver calls, beyond which evaluations carry negligible cost. The reader interested primarily in these results may proceed to~\cref{sec: Results}.

\section{Methods}\label{sec: methods}

\subsection{Setup and Notation}\label{sec: setup and notation}

We begin by defining the Hamiltonian of a physical model 
\begin{align}\label{eq: Hamiltonian}
    \mcH(\param)=\sum_q^Q p_q\mbf{H}_q
\end{align}
\noindent
where $\param=(p_1,\dots,p_Q)$ is a coefficient vector containing the Hamiltonian parameters and $\mbf{H}_q\in\reals^{d\times d}$ are constant, parameter-independent operators. Here, $d$ defines the dimension of the many-body Hilbert space. For example, a system of $N$ two-level sites has $d=2^{N}$. However, typically only a specific particle sector is relevant (for example, at half-filling) and $d$ does not represent the entire Hilbert space, but instead the particle-selected space, which still scales exponentially with the system size. We define the parameter-dependent ground state of this Hamiltonian to be $|\Psi(\param)\rangle\in\reals^{d}$ with ground state energy $E(\param)$.

Our primary goal is to obtain single-particle spectra in the form of a GF. For a given physical model, the fermionic GF is formally defined via the Lehmann representation of the time-dependent anticommutator $\mbf{G}_{ij}(\param,t)=\langle\Psi(\param)|\{\c_i(t),\cd_j\}|\Psi(\param)\rangle$, where $\c_i,\cd_j$ are particle creation and annihilation operators acting on sites $i$ and $j$, respectively. In frequency space, this results in a structure involving the resolvents of both the particle and hole excitations. 

To maintain a general notation agnostic to whether the operators in $\mbf{G}$ correspond to particle creation or annihilation, 
we adopt the following compact form:
\begin{align}\label{eq: GF}
\mbf{G}_\mbf{AB}(\param,z) = \bigg\langle\Psi(\param)\bigg|\mbfA
\frac{1}{
z\mbfI \pm (\mcH(\param)-E(\param)\mbfI)} \mbfB\bigg|\Psi(\param)\bigg\rangle
\end{align}

\noindent
where $z$ is a complex frequency (e.g., $i\omega_n$ for a Matsubara Green's function, or $\omega+i\eta$ for a real-frequency retarded Green's function with $\eta \to 0^+$). We use $\pm$ to describe the contribution from either particle addition ($-$) or removal ($+$). For brevity, we drop the diagonal shift $E(\param)$ in the resolvent in the following equations, and it should be assumed this shift is always included where $E(\param)$ is the eigen-energy of $|\Psi(\param)\rangle$.

\subsection{Reduced representation of the ground state via Eigenvector Continuation}\label{sec: eigenvector continuation}

Armed with the definitions given in~\cref{sec: setup and notation}, we begin to tackle the first requirement for evaluating~\cref{eq: GF}: determining the many-body ground state $|\Psi(\param)\rangle$ within a reduced subspace. As discussed in~\cref{sec: morec in brief}, this subspace supplies both the compact representation of the ground state as well as the directions for the parametric interpolation of the GF.

Given some set of parameter points $\bm{\param}_k = \{\param_1, \dots, \param_k\}$, we construct a low-energy subspace basis $\mcB = \left[|\Phi(\param_1)\rangle, \dots, |\Phi(\param_k)\rangle\right] \in \reals^{d \times k}$, built, e.g., from ground states of the Hamiltonian at these parameter points. This basis allows us to approximate $|\Psi(\param)\rangle$ at \textit{any} target parameter $\param^{\,*}$ by solving a $k \times k$ generalized eigenvalue problem

\begin{align}\label{eq: GEP}
    \mbf{h}(\param^{\,*})\varphi^{(0)} &= \lambda^{(0)}(\param^{\,*})\mbf{S}\varphi^{(0)}, \nonumber \\
    |\Psi(\param^{\,*})\rangle &\approx \mcB\varphi^{(0)}
\end{align}

\noindent
where $\mbf{h}(\param^*) = \sum_q^Q p_q^* \mcB^\dagger \mbf{H}_q \mcB$ is the reduced-basis Hamiltonian, $\mbf{S} = \mcB^\dagger \mcB$ is the overlap matrix, and $(\lambda^{(0)}(\param^*),\varphi^{(0)})$ are the ground state eigenvalue-eigenvector pair in the reduced space of order $k$. In Appendix~\ref{asec: adpative EC}, we detail how we use a greedy selection algorithm to determine the set $\bm{\param}_k$. 

Briefly, our approach is a two-step process. During an \emph{offline} phase, a minimal basis is first constructed before a scan over the parameter domain. This scan, which we refer to as the \emph{online enrichment} phase, appends additional vectors to the basis in parameter regions where an error surrogate -- which is entirely inferred within the reduced basis -- is above some threshold $\varepsilon$.

It should be noted that the vectors in $\mcB$ need not belong to the same family of Hamiltonian as $\mcH(\param)$. Methods such as non-orthogonal configuration interaction~\cite{Sundstrom2014,Oosterbaan2018} or EC for approximate ground states~\cite{agrawal2025EC,Rath2025,Atalar2024} demonstrate that $\mcB$ can be successfully constructed from states that are not necessarily the exact ground states of the model at $\param_i$. Consequently, we assume $|\Phi_i\rangle$ represents either exact or approximate ground states of the model of interest.

For physical models, the subspace used for EC is typically much smaller than the full Hilbert space ($k\ll d$) and has been shown to converge more rapidly than standard perturbation theory~\cite{sakar2021convergenceEC}. This ability to accurately represent the ground state manifold across phase transitions~\cite{Mejuto-Zaera2023quantumEC, agrawal2025EC} makes EC an ideal engine for constructing globally accurate ROMs. Finding the ``well-suited subspace'' $\mcB$ that spans these regimes is the most critical step in MOR+EC, and we turn to the next section to demonstrate its role in the ROM for the GF.

\subsection{Reduced-order models for evaluation of Green's functions by frequency interpolation}\label{sec: ROM by interp}

Following the prescription for finding the ground state in~\cref{sec: eigenvector continuation}, we seek to find a complementary ROM which describes the dynamical response of a many-body system to various driving frequencies, $z$, across a broad parameter regime $\param$. This task is a specific instance of parametric model order reduction (PMOR)~\cite{antoulas2020interpolatory, benner2016pmorsurvey}, which works as follows. We characterize the system by its transfer function $\gamma(\param, z)$, 

\begin{align}\label{eq: transfer functions}
    \gamma(\param, z) &= \mbf{a}(\param)^\top(z\mbfI \pm \mcH(\param))^{-1}\mbf{b}(\param)
\end{align}
\noindent
which is equivalent to $\mbf{G}_\mbf{AB}(\param, z)$ under the identification 
$\mbf{b}(\param)=\mbfB|\Psi(\param)\rangle$ 
and 
$\mbf{a}(\param)=\mbfA|\Psi(\param)\rangle$.
To circumvent the exponentially large dimension $d$ of the many-body Hilbert space, we construct a ROM of dimension $m \ll d$ that preserves the input-output mapping, or equivalently the transfer function $\gamma(\param,z)$.
We define the relationship between the full-order model and the ROM transfer functions as

\begin{align}\label{eq: reduced transfer functions}
    \gamma(\param, z) &\approx\widetilde{\gamma}(\param, z)
    =\widetilde{\mbf{a}}(\param)^\top(z\widetilde\mbfI \pm \widetilde\mcH(\param))^{-1}\widetilde{\mbf{b}}(\param) 
\end{align}
\noindent
where all $\widetilde{\Box}$ quantities are small, i.e., of order $m$.
The goal of PMOR is to construct a projection that ensures this approximation holds across the desired frequency and parameter ranges. Physically, this involves finding a subspace that captures both the system response (the resolvent's action at various $z$) and the parameter manifold (the variation of the states with $\param$).

A robust way to construct such a projection is through moment matching~\cite{antoulas2020interpolatory}. 
As a solution to the general problem of rational interpolation~\cite{Opmeer2012MORrationalinterp}, the $0^\text{th}$-order moments of $\gamma(\param,z)$ are evaluated at a selection of interpolation frequencies $\mbf{z}_r=\{z_i\}_{i=1}^r$ and parameters $\bm{\param}_k=\{\param_j\}_{j=1}^k$. The general framework also matches the derivatives of $\gamma(\param,z)$ at the cost of repeated application of the resolvent. Because $\mcB$ already supplies $k$ independent parameter directions at each $z_i$, we found that additional derivative directions are not necessary here.

\begin{figure}[t]
    \centering
    \includegraphics[width=0.9\linewidth]{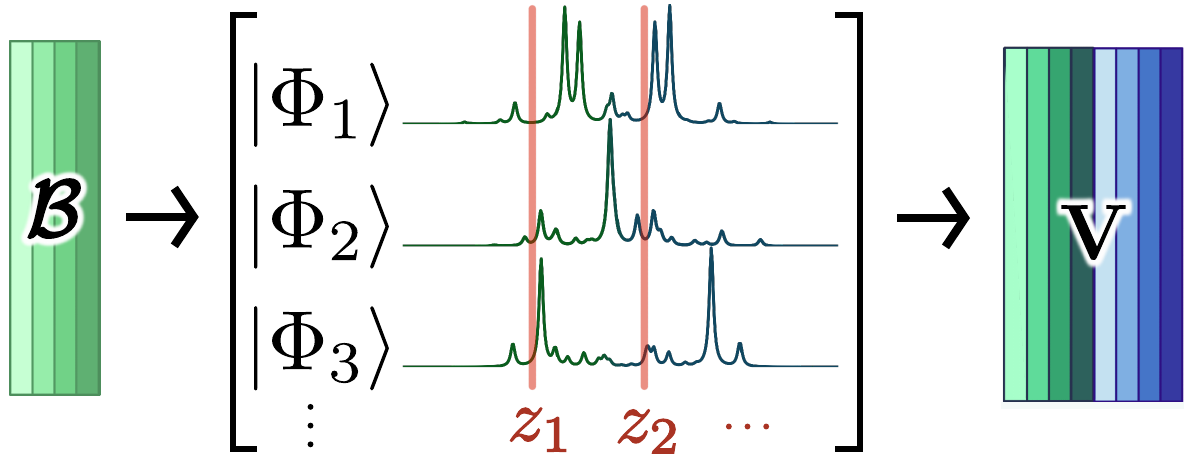}
    \caption{\textbf{Illustration of MOR+EC projections.} Once a basis $\mcB$ for EC is determined, a set of interpolation points $\mbf{z}_r=\{z_1,\dots,z_r\}$ are chosen to evaluate the columns of $\mbfV$ and $\mbfW$ (see~\cref{eq: PMOR projections}). Each state $|\Phi_i\rangle\in\mcB$ in principle has a unique spectrum.}
    \label{fig: MOREC projection}
    
\vspace{-6pt}
\end{figure}

We use $\mcB$ and $\mbf{z}_r$ to construct $\mbfV, \mbfW \in \complex^{d \times rk}$

\begin{align}\label{eq: PMOR projections}
\mbfV &= \left[v^{(1)}_1, \dots, v^{(k)}_1, \dots, v^{(1)}_r,\dots,v^{(k)}_r\right] \nonumber \\
\mbfW &= \left[w^{(1)}_1, \dots, w^{(k)}_1, \dots, w^{(1)}_r,\dots,w^{(k)}_r\right]
\end{align}

\noindent
where the vectors $v^{(j)}_i=(z_i\mbfI\pm\mcH(\param_j))^{-1}\mbf{b}(\param_j)$ and $(w^{(j)}_i)^\top = \mbf{a}(\param_j)^\top(z_i\mbfI\pm\mcH(\param_j))^{-1}$ represent ``snapshots" of the system $\gamma(\param,z)$. This process is depicted in~\cref{fig: MOREC projection}. We recall here that $d$ is an exponentially scaling particle-restricted space rather than the full Hilbert space. For example, when $\mbf{a}(\param_j)$ and $\mbf{b}(\param_j)$ are particle (hole) excitations to the ground state with $n$ occupied sites, then $d$ represents the $n+1$ ($n-1$) particle sectors, respectively.

By projecting the full-order model operators onto the subspace spanned by these snapshots, we obtain the reduced quantities

\begin{align}\label{eq: projected model}
\widetilde\mbfH_q &= \mbfW^\top\mbfH_q\mbfV, \quad \widetilde\mbfI = \mbfW^\top\mbfV, \nonumber \\
\widetilde{\mbf{b}}(\param) &= \mbfW^\top\mbf{b}(\param), \quad \widetilde{\mbf{a}}(\param)^\top = \mbf{a}(\param)^\top\mbfV
\end{align}

\noindent
where $(\cdot)^\top$ is generally the conjugate transpose for complex $z_i$.

The full Hamiltonian $\mcH(\param)$ can be efficiently
reconstructed through $\widetilde\mcH(\param)=\sum_q^Qp_q\widetilde\mbfH_q$. Crucially, once the $Q$ terms of $\widetilde\mbfH_q$ are pre-computed ``once-and-for-all," the ROM transfer function $\tilde\gamma(\param,z)$ in~\cref{eq: reduced transfer functions} can be evaluated for any Hamiltonian parameter vector $\param$ and frequency $z$ at a negligible computational cost relative to the full-space calculation. The cost of reconstructing $\widetilde\mcH(\param)$ and evaluating the ground state scales typically cubically with the number of orbitals in the system, i.e. it is equivalent to a mean-field calculation~\cite{Rath2025,Giuli2026}. Then, the evaluation of the GF scales polynomially with $m$. 

\subsection{Combining Model Order Reduction (MOR) with Eigenvector Continuation (EC)}\label{sec: MOREC}

A central challenge in PMOR is selecting parameter samples $\param$ and interpolation points $z$ that yield projection spaces rich enough to capture the system's physics. Standard strategies, such as greedy sampling based on error indicators~\cite{bui-thanh2008greedyMOR} or sparse grid constructions~\cite{Bungartz_Griebel_2004_sparsegrids}, can be computationally taxing due to repeated high-fidelity solves and may scale poorly with parameter dimension. 

To address this, we combine the adaptive sampling of EC with the interpolatory framework of PMOR, decoupling the task of finding support across $\param$ and the frequency variable at once. Specifically, we use the adaptive basis selection method detailed in Appendix~\ref{asec: adpative EC} to first determine the physical subspace $\mcB$, and then build the projection matrices $\mbfV$ and $\mbfW$ using these states as the directions for our interpolation points $\mbf{z}$ (see~\cref{fig: MOREC projection}). We refer to this framework as MOR+EC; it functions as a two-level solver where the ground state is first resolved in the basis $\mcB$ of order $k$, followed by the solution of the GF in the basis of $\mbfV$ and $\mbfW$ of order $m\sim kr\ll d$, where we recall that $r$ is the number of frequency interpolation points, and $m\sim kr$ accounts for $m<kr$ after a rank-revealing orthonormalization of $\mbfV,\mbfW$.

\begin{figure}
    \centering
    \includegraphics[width=0.9\linewidth]{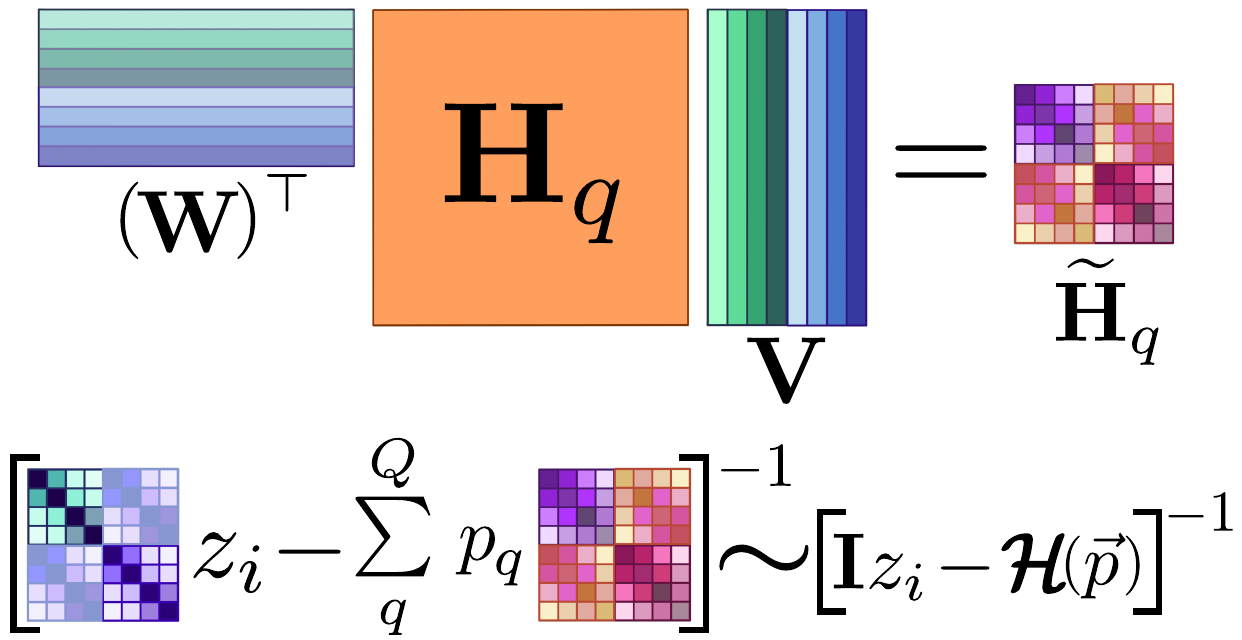}
    \caption{\textbf{Projected quantities for the reduced-order model using MOR+EC.} Given a full order model $\mcH(\param)=\sum_q^Qp_q\mbfH_q$ and projections built with MOR+EC, the resolvent of the full order model is approximated using the projected quantities $\widetilde\mbfI=(\mbfW)^\top\mbfV$ (multiplying $z_i$) and $\widetilde\mbfH_q=(\mbfW)^\top\mbfH_q\mbfV$.}
    \label{fig: MOREC projected quantities}
    
\vspace{-6pt}
\end{figure}

Concretely, the first level of the solver finds the ground state within the EC basis $\mcB$, and the resulting reduced-space eigenvalue and eigenvector pair ($\lambda^{(0)}(\param),\varphi^{(0)}$) are passed as inputs to the ROM. For a given parameter $\param$, $\widetilde{\mbf{a}}(\param)$ and $\widetilde{\mbf{b}}(\param)$ are constructed using $\varphi^{(0)}$, and the energy shift to the resolvent becomes $\lambda^{(0)}$. The form of the particle-addition contribution to the resolvent is shown in~\cref{fig: MOREC projected quantities}, where we note that the energy shift $\lambda^{(0)}$ is absorbed into $\widetilde\mcH(\param)=\sum_q^Qp_q\mbfH_q$.

We note that, in general, computing the snapshot vectors within~\cref{eq: PMOR projections} requires factorizations of the resolvent $\mbf{R}(\param_j,z_i)=(z_i\mbfI-\mcH(\param_j))^{-1}$ for every $z_i$ and $\param_j$. However, in our case, we opt for a parameter-independent resolvent $\mbf{R}_0(z_i)=(z_i\mbfI-\mcH_0)^{-1}$ with respect to a reference Hamiltonian $\mcH_0=\mcH(\param^{\,0})$ with $\param^{\,0}$ centered at the parameter domain of interest to reduce the required factorizations from $kr$ to $r$. This reduces the cost of constructing the ROM significantly, due to the fact that in most cases $k\gg r$. We address the details of this approximation and its consequences in Appendix~\ref{asec: param indep resolvent}.

\subsection{Application: Dynamical Mean Field Theory}\label{sec: DMFT}

To demonstrate the robustness of MOR+EC for computing low-order GFs for parameter space exploration, we apply our methods in the Dynamical Mean Field Theory (DMFT) framework~\cite{Kotliar1996,Kotliar2006,Zgid2011,Paul2019}. DMFT maps a fully correlated model, which cannot be directly solved accurately, to a simpler Anderson impurity model (AIM), which is more computationally tractable. This model has one, in the case of the single-impurity Anderson model (SIAM), or a few correlated spinful orbitals, each coupled to a noninteracting electron bath. The bath is described by hoppings from the impurity orbital $i$ to the bath orbitals $b$ ($V^i_b$) and on-site potential energies ($\epsilon^i_b$). 

Its Hamiltonian for $N_I$ impurity orbitals and $N_B$ bath orbitals per impurity orbital is as follows:

\begin{align}\label{eq: impurity H}
    \mcH^\text{imp}&=\mbfH^\text{loc}\nonumber \\
    &+\sum_{i,b}^{N_I,N_B} \sum_\sigma V^i_b(\hat{d}^\dagger_{i\sigma}\hat{c}^\phant_{ib\sigma} + \text{h.c.}) + \epsilon^i_b \hat{c}^\dagger_{ib\sigma} \hat{c}^\phant_{ib\sigma} 
\end{align}

\noindent
where $\hat{d}^{(\dagger)}_{i\sigma}$ annihilates (creates) a particle on impurity $i$ with spin $\sigma$ and $\hat{c}^{(\dagger)}_{ib\sigma}$ annihilates (creates) a particle on bath orbital $b$ with spin $\sigma$ coupled to impurity orbital $i$. The impurity-only terms in $\mbf{H}^\text{loc}$ depend on the number of impurity orbitals and the structure of the correlated model of interest. In this work, we apply our methods and perform DMFT calculations on the half-filled single-band and two-band Hubbard model on a Bethe lattice. 

The local part for the single-band model is

\begin{align}\label{eq: 1B local H}
    \mbfH^\text{loc}_\text{1B}=U\hat{n}_{1\uparrow}\hat{n}_{1\downarrow} + \nu_1\sum_\sigma \hat{n}_{1\sigma}
\end{align}

\noindent
with $\hat{n}_{i\sigma}=\hat{d}^\dagger_{i\sigma}\hat{d}^\phant_{i\sigma}$ and for the two-band model, 

\begin{align}\label{eq: 2B local H}
    \mbfH^\text{loc}_\text{2B} &=U \sum_{i=1}^{2} \hat{n}_{i\uparrow} \hat{n}_{i\downarrow} + U'\sum_{i\neq j,\sigma\sigma'}^{2} \hat{n}_{i\sigma} \hat{n}_{j\sigma'} +\sum_{i\sigma}^{2} \nu_{i} \hat{n}_{i\sigma} \nonumber \\
    &+ J\sum_{i\neq j,\sigma\neq\sigma'}^{2} \hat{d}^{\dagger}_{i\sigma} \hat{d}^\dagger_{j\sigma'} \hat{d}^\phant_{i\sigma} \hat{d}^\phant_{j\sigma'} + J \sum_{i\neq j}^{2} \hat{d}^\dagger_{i\uparrow} \hat{d}^\dagger_{i\downarrow} \hat{d}^\phant_{j\uparrow} \hat{d}^\phant_{j\downarrow}. 
\end{align}

\begin{figure}
    \centering
    \includegraphics[width=0.9\linewidth]{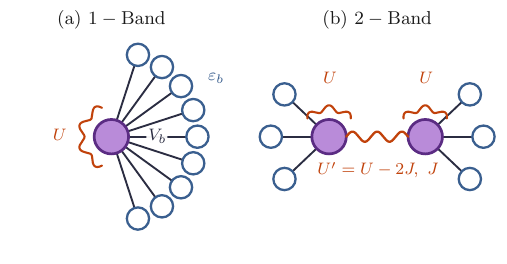}
    \caption{\textbf{Impurity models used in this work.} (a) The single-impurity model involves a single interacting orbital (filled circle) with interactions (squiggly lines) described by~\cref{eq: 1B local H} coupled to 9 noninteracting bath orbitals (empty circles). (b) The two-band model involves two interacting orbitals with intra- ($U$) and inter- ($U'$) orbital interactions described by~\cref{eq: 2B local H}. Each orbital is coupled to 3 noninteracting orbitals. }
    \label{fig: models}
    
\vspace{-6pt}
\end{figure}

\noindent
Here, $U$ and $U'=U-2J$ are the intra- and inter- Coulomb repulsion between spins on impurity orbitals $i$ and $j$. $\nu_{i}$ is the one-body potential for the impurities (corresponding to the on-site chemical potential) and $J$ is the Coulomb exchange strength that governs Hund's rules. These models are depicted in~\cref{fig: models}.

This impurity model is related back to the lattice model self-consistetly through the local self-energy; if the impurity model self-energy reproduces the local Green's function of the lattice model, the embedding description is considered self-consistent and thus the DMFT is converged. Details about the DMFT self-consistency procedure are summarized in Appendix~\ref{asec: DMFT details}.

While DMFT has been effective at capturing the correlated behavior of real materials~\cite{Kotliar2006,Paul2019}, the computational expense of evaluating the impurity GF $\mbf{G}^\text{imp}$ after updating the impurity model's parameters at every iteration of DMFT can be prohibitive for complex systems. 

Here is where MOR+EC can offer an advantage by finding a reduced representation of $\mbf{G}^\text{imp}$ that is compatible with all relevant Hamiltonian parameters using the methods described in the previous sections. Once an accurate ROM is determined by building the projections in~\cref{eq: PMOR projections}, the projected Hamiltonian terms $\widetilde\mbfH_q$ are computed \textit{once-and-for-all}, then each parameter instance $\param$ requires finding the ground state of the Hamiltonian in the reduced basis $\mathbf{h}(\param)$ of order $k$ then taking the resolvent of $\widetilde\mcH(\param)$ of order $m$. 

For our results, we use $N_B=9$ for the single-band model and $N_B=3$ per impurity orbital in the two-band model, resulting in 10 and 8 total spinful orbitals, respectively. This corresponds to full-order dimensions of size $d_\mathrm{GS}=63504$ for the half-filled ground state and $d_{p/h}=52920$ for the particle/hole part of the GF for the single-band impurity model. Likewise, for the two band model, $d_\mathrm{GS}=4900$ and $d_{p/h}=3920$.

\section{Results}\label{sec: Results}

\subsection{Worked example: the SIAM}

Before discussing our DMFT application, we present a simple example to illustrate the intuition behind MOR+EC. In this example, we aim to build a ROM using MOR+EC that can accurately capture the spectrum of a single-band impurity model with 5 bath sites as we vary the interaction strength, $U$, and keep the bath parameters fixed.

Using the offline setting for basis selection (see Appendix~\ref{asec: adpative EC}), we find a reduced basis $\mcB$ of order $k=12$ ($3\%$ of the half-filled particle sector dimension $d_\mathrm{GS}=400$) such that we achieve near-perfect accuracy in representing the ground state over 200 test instances ($\ge99.95\%$ fidelity with the true ground state). Then, using this basis as our directional interpolation, we employed the widely-used IRKA algorithm~\cite{gugercin2008H2modelreduction,antoulas2020interpolatory,FLAGG2012688} for greedily selecting $r=6$ complex interpolation frequencies.

\begin{figure}
    \centering
    \includegraphics[width=0.95\linewidth]{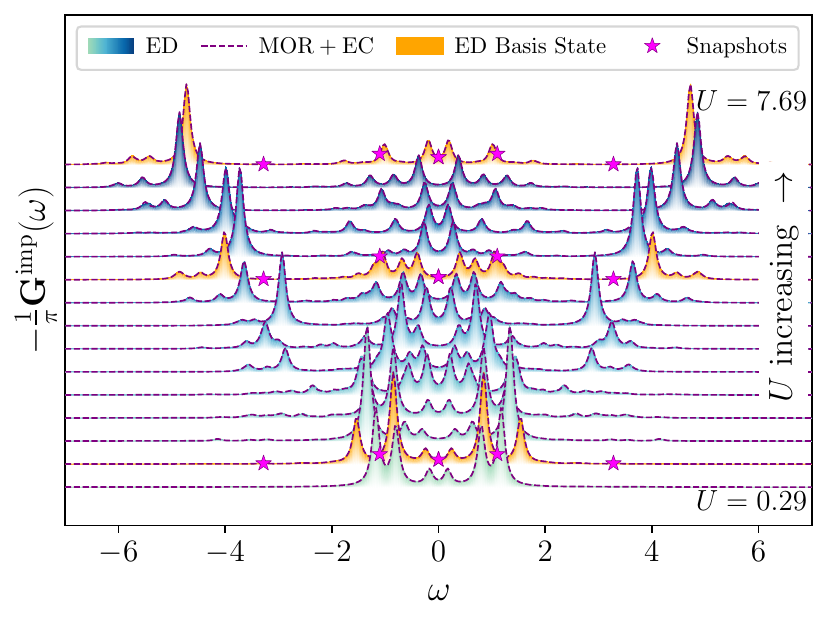}
    \caption{\textbf{Simple example of MOR+EC in action.} Using the SIAM with 5 bath sites, we benchmark MOR+EC's performance when only one parameter is varied (fixed bath, varying $U$) and show that MOR+EC completely captures the entire spectrum of states outside the EC basis $\mcB$ and at $\omega$ values outside of our interpolation set $\mbf{z}_r$. The magenta stars show what snapshots are in the MOR+EC projections (\cref{eq: PMOR projections}) and all other data is found with the ROM.}
    \label{fig: toy example}
    
\vspace{-6pt}
\end{figure}

After building this ROM, we sampled a few interaction strengths $U$ to obtain the results in~\cref{fig: toy example}. Here, we show the single-band impurity model's density of states computed for 500 frequency points using the ROM. We show ED results as a shaded region, and these are perfectly enveloped by the MOR+EC data in the dashed purple line.

We visually demonstrate where the snapshot data in the ROM is obtained with the magenta stars. For illustrative purposes, we plot the spectra from a few sample EC basis states (highlighted in orange) with the stars marking the $r=6$ frequency samples (two of the six lie on the imaginary axis and coincide at $\omega=0$). The stars represent the snapshots where full-order calculations were performed to create the projections described by~\cref{eq: PMOR projections}. All other data points comprising the dashed purple line (labeled MOR+EC) are obtained solely in the ROM, which after a rank-revealing QR, ended up being of order $m=60$ ($\sim$80\% space reduction from the $d_{p/h}=300$ particle/hole removal sectors). The absolute relative error of the worst data point across all test cases was $\sim10^{-2}$.

All details related to our numerical results may be found in Appendix~\ref{asec: numerical details}.

\begin{figure*}[t]
    \centering
    \includegraphics[width=\linewidth]{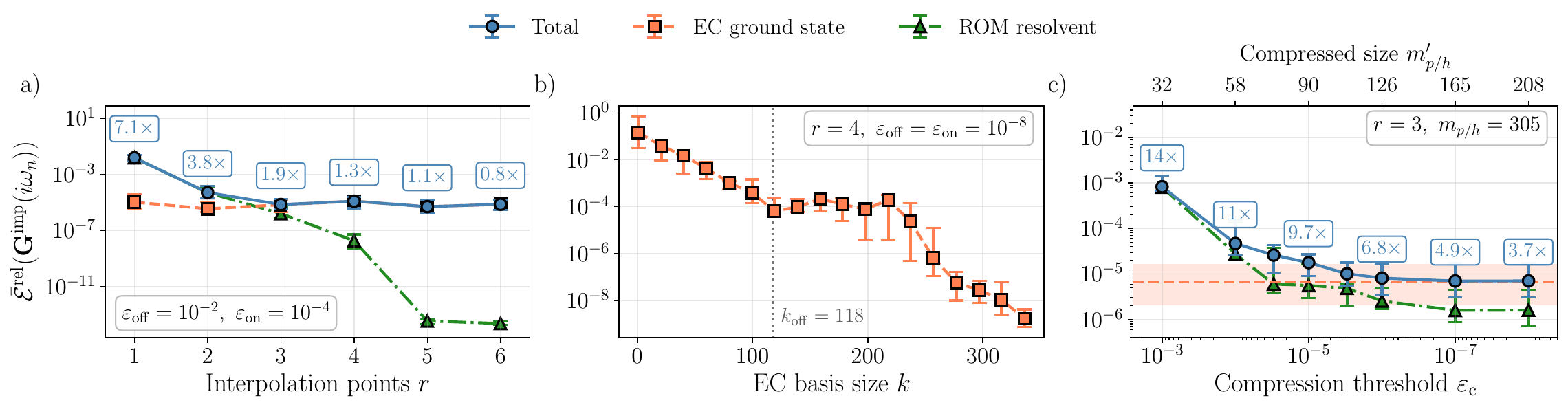}
    \caption{\textbf{Average relative error in DMFT-converged impurity GFs across 100 two-band instances.} (a) Total error (blue circles, solid line) split into the EC channel (orange squares, dashed line, exact resolvent in basis $\mcB$) and ROM resolvent channel (green triangles, dot-dashed line). Beyond $r=3$, error hits the EC floor. Annotations show median impurity solver speedup (ED vs. MOR+EC). (b) Lowering the variance threshold to $10^{-8}$ ($k_\off=118$ for the offline setting, $k=336$ after online enrichment) reduces the EC floor at the cost of a larger ROM. (c) The $r=3$ ROM of panel (a) after the compression of Appendix~\ref{asec: compression}, against the threshold $\varepsilon_\mathrm{c}$. The upper axis gives the compressed size, and the annotations are the speedup. The compression leaves the EC ground state unchanged, so its channel is a line, with a shaded band for its 25$^\text{th}$--75$^\text{th}$ percentile range (the whiskers of the other points). Speedups and ROM sizes are summarized in~\cref{tab:gf_error_channels_timing}.}
    \label{fig: MOREC error}
    
\vspace{-6pt}
\end{figure*}

\subsection{Accuracy and Cost}

A measure we use for the accuracy of the presented quantities is the absolute relative error, defined as

\begin{align}
    \relerror(f(x))=\frac{1}{N_x}\sum_x\frac{\|f_\text{MOR+EC}(x)-f_\text{ED}(x)\|}{\|f_\text{ED}(x)\|}
\end{align}

\noindent
where $f(x)$ is either a number or a matrix-valued quantity. The brackets $\|\cdot\|$ denote either the absolute value for the former and the 2-norm for the latter.

\cref{fig: MOREC error} shows the average relative error of the DMFT-converged impurity GF for the two-band model with $J=0$, pooled over 100 DMFT instances initiated at various $U$ and $t_2/t_1$ settings. To isolate the error contributed by MOR+EC alone, we take its converged fixed point and re-run ED at that same fixed point. The result splits the error contributions in two: past a small number of interpolation points, further accuracy comes from growing the EC basis, not the ROM.

Panel (a) fixes $\varepsilon_\off=10^{-2}$, $\varepsilon_\on=10^{-4}$, two convergenge thresholds used to select the basis $\mcB$ (see Appendix~\ref{asec: adpative EC} for details), and sweeps the number of interpolation frequencies $r$. The offline EC basis size is $k_\off=16$ for every instance, and online enrichment then brings it to $k\approx102$--$110$. At $r=1,2$ the total error tracks above the EC-only curve, meaning the ROM's resolvent is still the limiting factor; by $r=3$ the error has caught up to the EC floor at $\sim10^{-5}$, and adding more interpolation points past that only inflates the ROM -- shrinking the solver speedup (annotated above each point) -- without buying any further accuracy.

Panel (b) shows that floor is not fundamental: tightening both thresholds to $\varepsilon_\off=\varepsilon_\on=10^{-8}$ (with $r=4$, chosen so the ROM resolvent channel stays below the EC channel throughout) grows the offline basis to $k_\off=118$ and, after online enrichment, to $k=336$. At this basis size, the EC-limited error floor is dropped to $\sim10^{-9}$.

\begin{table}[tb]
  \centering
  \setlength{\tabcolsep}{4pt}
  \footnotesize
  \begin{tabular}{@{}llrrrrrr@{}}
    \toprule
     & & \multicolumn{3}{c}{Uncompressed} & \multicolumn{3}{c}{Compressed} \\
    \cmidrule(lr){3-5}\cmidrule(lr){6-8}
    Panel & $r$ & $m_{p/h}$ & GF (ms) & total & $m'_{p/h}$ & GF (ms) & total \\
    \midrule
    (a) & 1 & 108 & 5.94 & $7.1\times$ & 86 & 4.55 & $8.9\times$ \\
    (a) & 2 & 220 & 11.34 & $3.8\times$ & 90 & 4.44 & $9.3\times$ \\
    \textbf{(a)} & \textbf{3} & \textbf{305} & \textbf{22.91} & $\boldsymbol{1.9\times}$ & \textbf{90} & \textbf{4.32} & $\boldsymbol{9.5\times}$ \\
    (a) & 4 & 374 & 33.40 & $1.3\times$ & 91 & 4.26 & $9.7\times$ \\
    (a) & 5 & 430 & 40.86 & $1.1\times$ & 91 & 4.26 & $9.7\times$ \\
    (a) & 6 & 502 & 53.70 & $0.8\times$ & 90 & 4.29 & $9.6\times$ \\
    \midrule
    (b) & 4 & 1098 & 245.82 & $0.2\times$ & 90 & 4.66 & $5.2\times$ \\
    \bottomrule
  \end{tabular}
  \caption{\textbf{Per-call impurity-solver cost of MOR+EC for the two-band model, pooled over 100 parameter points of the DMFT scan.} $m_{p/h}$ is the mean order of the particle and hole contributions to the GF and $m'_{p/h}$ its value after the compression of Appendix~\ref{asec: compression} with $\varepsilon_\mathrm{c}=10^{-5}$. ``Total'' is the median per-point speedup of the full solver call against ED on the same baths (48\,ms ground state, 15\,ms GF), with all solvers timed in turn on the same compute node. The ground state solve of MOR+EC takes 0.5\,ms in rows (a) and 4.2\,ms in row (b) and is unchanged by the compression. Rows (a) and (b) are the configurations of Fig.~\ref{fig: MOREC error}; row (b) is its last data point ($k=336$). The bold line is the ROM used in~\cref{sec: 2B results}.}
  \label{tab:gf_error_channels_timing}
\vspace{-6pt}
\end{table}

This trade-off comes from the choice to couple EC and ROM growth by default: for every vector appended to $\mcB$, we append $r$ columns to the projections that build the ROM (\cref{sec: MOREC}), so pushing the error floor down with a larger EC basis also inflates $m$, and \cref{tab:gf_error_channels_timing} shows that cost directly. Panel (b)'s tighter thresholds buy more than three orders of magnitude of accuracy but leave the GF evaluation slower than ED's ($0.1\times$). That coupling between the growth in $\mcB$ and the ROM is a choice, not a requirement. 

Within the space spanned by the ROM basis, the $m\sim kr$ directions would be flexible enough to describe the wider family of impurity GFs visited along the DMFT iterations, but we are only interested in those that are compatible with the DMFT self-consistency of the model in question. We therefore compress the ROM after its construction by a proper orthogonal decomposition~\cite{willcox2002balancedpod,benner2016pmorsurvey,bonilla2022training,duget2024EC} of its own solution vectors at the converged points of the coarse scan (Appendix~\ref{asec: compression}), keeping the directions above a relative threshold $\varepsilon_\mathrm{c}$.

Panel (c) applies this compression to the $r=3$ ROM of panel (a). Raising the threshold shrinks the ROM and increases the speedup (annotations), at first with no cost to accuracy. The total error of the ROM added by the compression sits below the EC floor (orange line) until the compression brings the ROM below half its original size of $m_{p/h}=305$, and only then is the total error raised. At this step, the compressed size is set by $\varepsilon_\mathrm{c}$ and not $r$; the $r=6$ ROM from panel (a) compresses to the same sizes as the $r=3$ ROM at each $\varepsilon_\mathrm{c}$, meaning $r$ only matters for accuracy during enrichment, not speedup. We use $\varepsilon_\mathrm{c}=10^{-5}$ for both the single-band and two-band models and the numbers reported in~\cref{tab:gf_error_channels_timing}.

\subsection{Computational Speedup}\label{sec: speedup}

\subsubsection{Doping the single-band Hubbard Model}\label{sec: 1B results}

Our first benchmark is DMFT self-consistency for the single-band model, tracked through the impurity occupation $n_\text{imp}=\langle\Psi|(\hat{n}_{1\uparrow}+\hat{n}_{1\downarrow})|\Psi\rangle$ as the chemical potential $\nu_1:=\mu$ is varied. We build the EC basis $\mcB$ for the ROM using a coarse scan, setting $\varepsilon_\off=10^{-2}$, $\varepsilon_\on=10^{-1}$, over nine values of $U/W$ from $2.5\times10^{-3}$ to $2.0$ and eight of $\mu/W$ from $-1$ to $0.75$, i.e.\ 72 self-consistent DMFT scans. This coarse scan's sole purpose is online enrichment of the basis $\mcB$, which results in a basis of size $k=132$ (a 99.79\% reduction in $d_\mathrm{GS}$) and $m_{p/h}=392$, compressed to $m'_{p/h}=203$ with $\varepsilon_\mathrm{c}=10^{-5}$ (99.62\% in $d_{p/h}$).

\begin{figure}[t]
    \centering
    \includegraphics[width=0.95\linewidth]{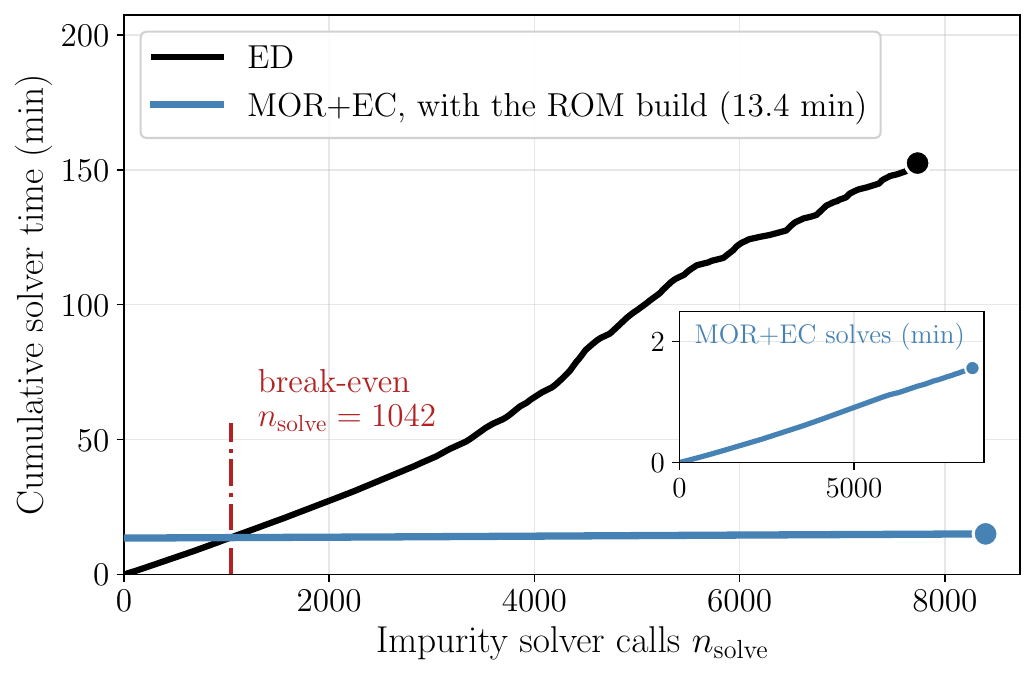}
    \caption{\textbf{Break-even point for the reduced-order model payoff for the single-band model.} After $\sim$1040 calls to an ED-based impurity solver, the cost to build the ROM with MOR+EC is paid off. (inset) All subsequent calls to the ROM impurity solver evaluate with a negligible cost of 1.6 minutes for 8397 calls to the ROM solver versus 2.5 hours for 7734 calls to the ED solver.}
    \label{fig: N10 break even err}
\vspace{-6pt}
\end{figure}

That ROM is then deployed for a fine-grained scan of 957 DMFT self-consistency loops in the same parameter domain (more than the 72 used to build it, and with more total impurity-solver calls than ED needs for the same grid, since GF error occasionally shifts MOR+EC to a different fixed point). \cref{fig: N10 break even err} tracks the cumulative solver time for both solvers across every call in that scan, ROM construction included. ED accumulates cost linearly, MOR+EC's one-time build cost (13.4 minutes) is repaid after $\sim$1040 ED-equivalent calls, and everything after that is nearly free -- 1.6 minutes total for MOR+EC's 8397 calls against 2.5 hours for ED's 7734.

\begin{figure}[t]
    \centering
    \includegraphics[width=0.95\linewidth]{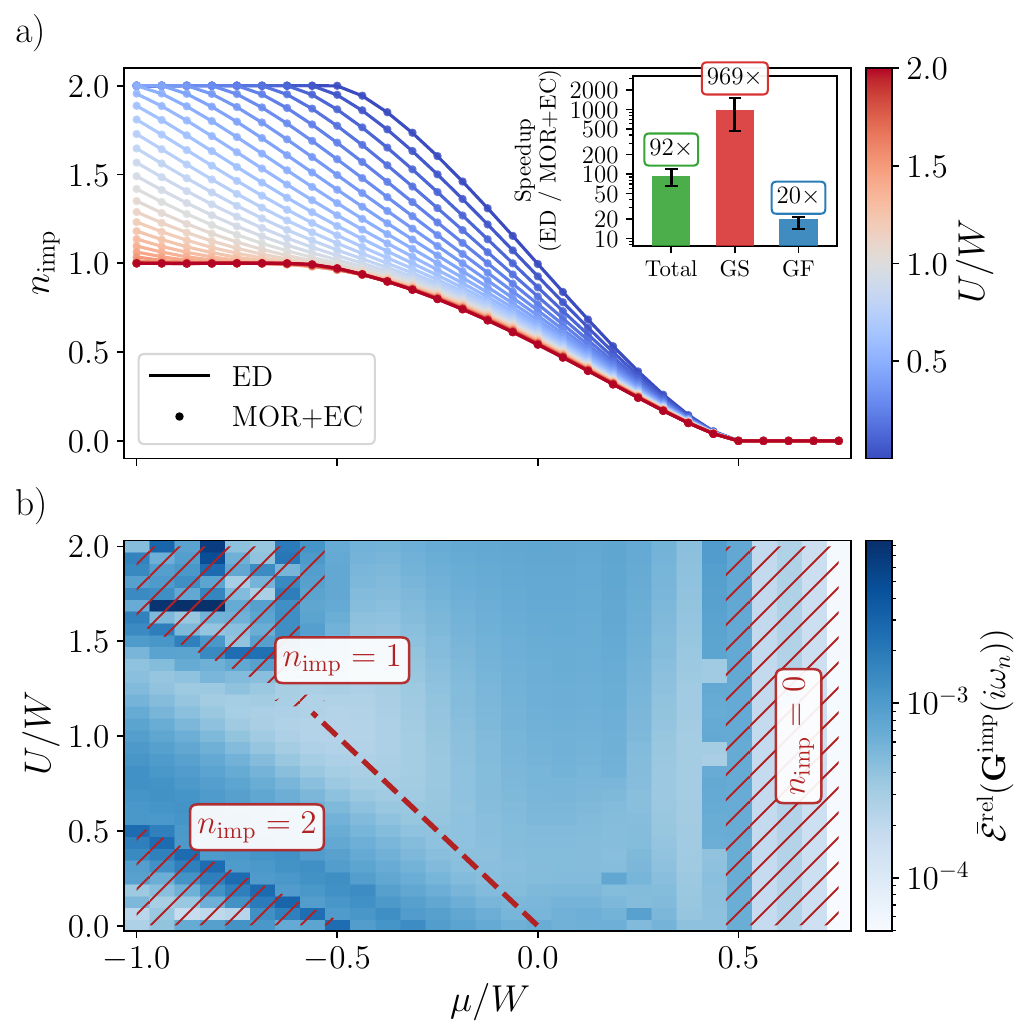}
    \caption{\textbf{Impurity occupation and GF error as the chemical potential $\bm{\mu/W}$ is adjusted.} (a) For the single-band Hubbard model, fine-grained DMFT scans of the chemical potential $\mu$ for different interaction strengths $U/W$ reveal a plateau at $n_\text{imp}=1$ for strong interactions. (Inset) Median wall-time speedup MOR+EC offered for each call to the impurity solver on a log scale, in terms of total solver time (green, Total) then separated by ground state solve (red, GS) and impurity GF solve (blue, GF). The median is annotated above each bar. Whiskers represent the $5^\mathrm{th}$ and $95^\mathrm{th}$ percentiles of the speedup data. (b) The relative error profile of the GF for the same data as (a), with integer filling denoted by the hatched areas and the dashed line (corresponding to $n_\mathrm{imp}=1$).}
    \label{fig: occ v mu}
    
\vspace{-6pt}
\end{figure}

\cref{fig: occ v mu}(a) shows what that scan actually resolves: at weak interactions, a negative chemical potential drives double occupation ($n_\text{imp}=2$). However, as $U$ grows, the system instead favors single occupation, visible as a plateau at $n_\text{imp}=1$. ED and MOR+EC agree closely throughout, with a median speedup of $969\times$ for the ground state and $20\times$ for the GF resulting in a total solver speed up of $92\times$ (inset), which is consistent with the wall-clock numbers already given above.

\begin{figure*}[t]
    \centering
    \includegraphics[width=0.9\linewidth]{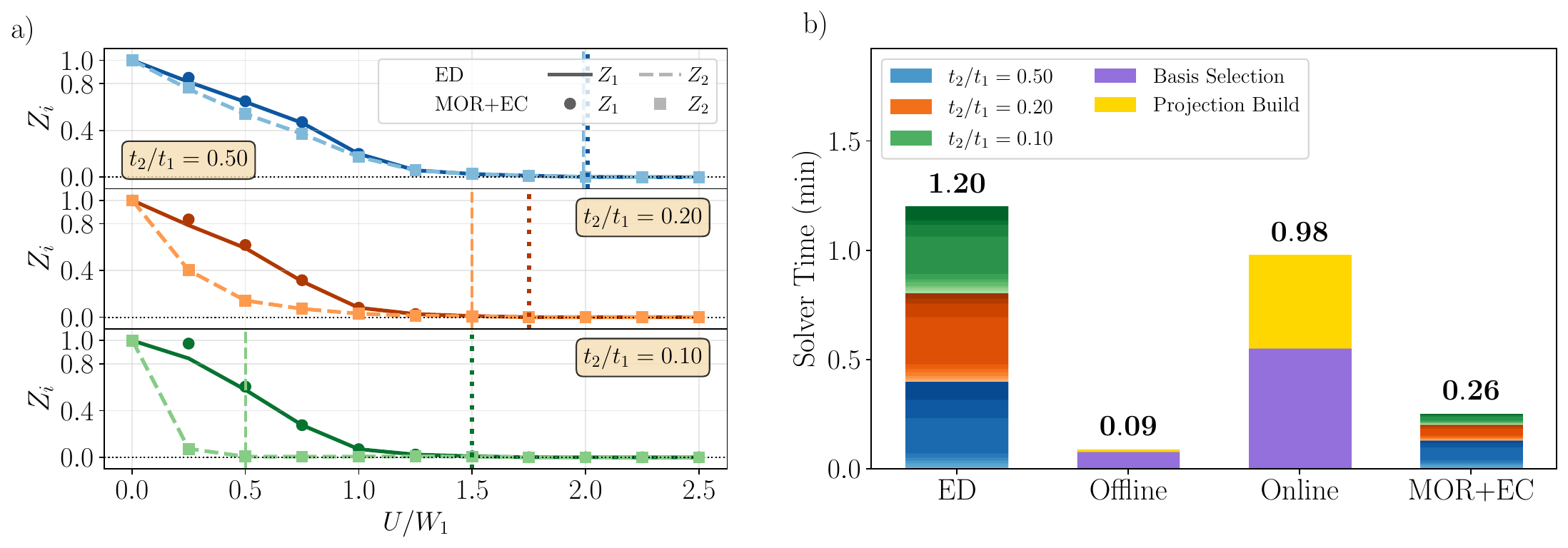}
    \caption{\textbf{Comparison of evaluation time and value of quasiparticle weights for the two-band model with $\bm{J=0}$ at different ratios $\bm{t_2/t_1}$.} (a) The dashed vertical lines indicate where $Z_i$ -- calculated with MOR+EC -- drops below a threshold of $10^{-2}$. (b) The wall time for all calls to the impurity solver for the DMFT scans in (a) is compared, as well as the offline and online MOR+EC build time. The color gradient for the ED and MOR+EC bars corresponds to the total impurity solver time during DMFT self-consistency at different interaction strengths $U$, with lighter being weaker and darker being stronger.}
    \label{fig: OSMT 2B}
    
\vspace{-6pt}
\end{figure*}

\cref{fig: occ v mu}(b) shows the average error of the GF across all solver evaluations per DMFT scan is small, lowest where the impurity is unoccupied, half-filled, or doubly-occupied (hatched regions), and highest on the boundaries between those regions. Note that in the present case, this error is not purely a ROM artifact. ED and MOR+EC can converge to slightly different self-consistent baths, part of what is shown in~\cref{fig: occ v mu} is fixed-point drift rather than the intrinsic resolvent error isolated in \cref{fig: MOREC error}.

That boundary sensitivity is also where the solver speedup is largest, which can be explained for both the ground state and GF cases. The case for the ground state is clear; ED finds it with Lanczos, whose iteration count is set by how well-separated the lowest eigenvalue is from the rest of the spectrum. Where that gap narrows, ED needs more iterations and slows down. MOR+EC's ground state, by contrast, is not found iteratively. It solves a generalized eigenvalue problem of fixed size $k=132$ in the EC basis with a direct, dense eigensolver, so its cost does not depend on how close two eigenvalues happen to be. As a result, MOR+EC's ground state solve time stays within 7\% of its median over the points timed for the inset of~\cref{fig: occ v mu}(a), regardless of $U$ or $\mu$, whereas ED's varies by a factor of 3.3 between its 5$^\text{th}$ and 95$^\text{th}$ percentiles, which is the spread of the ground-state whiskers.

The case for the GF evaluation is different, and explains why the whiskers in~\cref{fig: occ v mu}(a) are narrower here. Both solvers evaluate the GF with a Lanczos-type continued-fraction method, so neither is fully immune to parameter dependence. We find that ED's GF time varies by a factor of 2.5 between the same percentiles, and MOR+EC's by 1.9. What keeps MOR+EC fast is that its Lanczos recursion runs inside the reduced space $m$ rather than the particle-selected Hilbert space ED works in, so even comparable relative variation leaves its absolute cost an order of magnitude smaller throughout.

\subsubsection{The orbital-selective Mott transition in the two-band Hubbard Model}\label{sec: 2B results}

We now repeat this comparison for the two-band case, using the orbital-resolved quasiparticle weight $Z_i=(1-\frac{\mathrm{Im}[\mathbf{\Sigma}_{ii}(i\omega_n)]}{\omega_n}|_{\omega_n\to 0})^{-1}$ at bandwidth ratio $t_2/t_1$ (with $t_1=1$) both as an accuracy benchmark for the GF, for which $Z_i$ is highly sensitive, and as the order parameter for the transition itself. $Z_i\to1$ characterizes a conducting phase, while $Z_i\to0$ is insulating. At $J=0$, decreasing $t_2/t_1$ drives an orbital-selective Mott transition (OSMT), resulting in the narrow band's $Z_2$ collapsing at weaker $U/W_1$ than the wide band's $Z_1$~\cite{Ferrero20052Bdmft,Inaba20052Bdmft, deMedici2005osmt}, so the narrow band turns insulating while the wide band stays conducting.

\cref{fig: OSMT 2B} shows MOR+EC's accuracy while the ROM is still under construction: an offline pass gives $k_\off=16$, then online enrichment ($r=3$, $\varepsilon_\on=10^{-4}$) proceeds through a coarse scan of 55 points across $U/W_1$ and $t_2/t_1$. Panel (a) shows three representative $t_2/t_1$ slices from that coarse scan. Early on ($t_2/t_1=0.10$, bottom), the basis is still incomplete and ED/MOR+EC disagree visibly; by the last points ($t_2/t_1=0.50$, top), the ROM is nearly complete and the two agree closely. Panel (b) shows that even while paying for this construction, MOR+EC's total time for the coarse scan (1.33 minutes -- 0.09 offline, 0.98 online, 0.26 solver) is comparable to ED's impurity-solver time alone (1.20 minutes); most of that 1.33 minutes is online enrichment, not the solver.
The resulting ROM -- $k=108$ for $\mcB$, $m_{p/h}=305$ per orbital, compressed to $m'_{p/h}\approx90$ -- gives a $\sim$97.80\% reduction in the ground state space and $\sim$97.70\% in the GF space.

\begin{figure*}[t]
    \centering
    \includegraphics[width=0.9\linewidth]{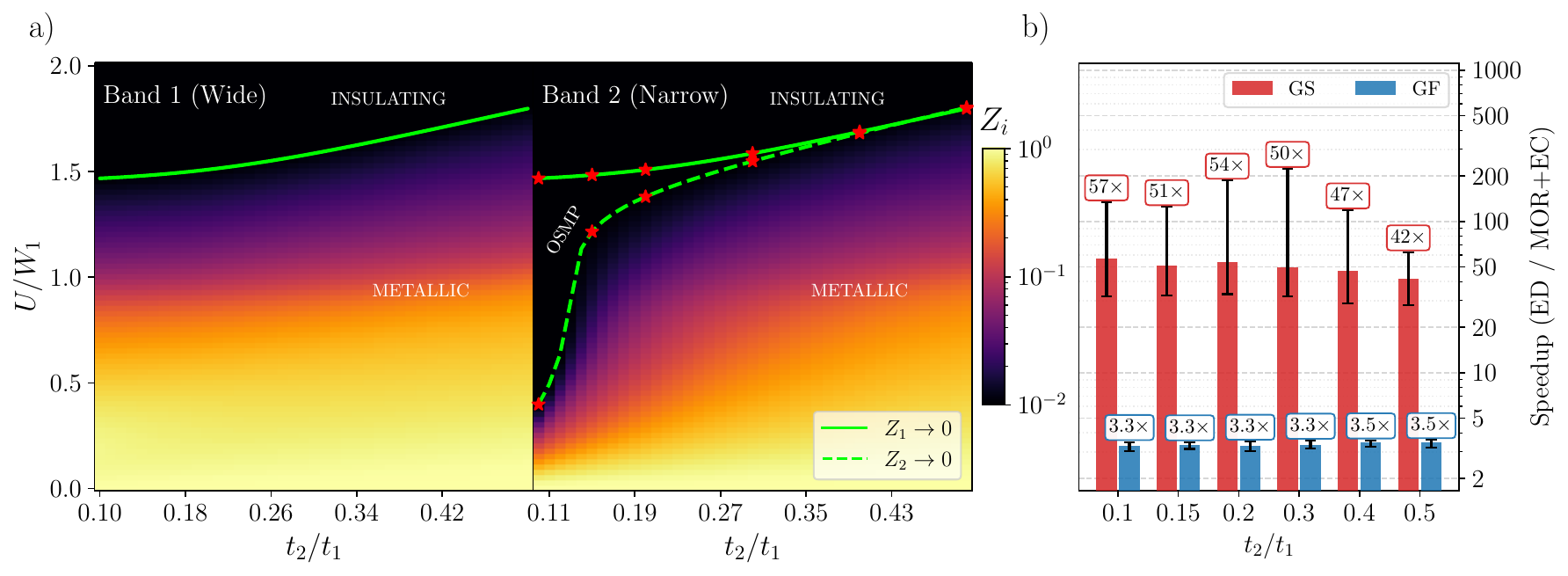}
    \caption{\textbf{Fine-grained phase diagram for the two-band model at $\bm{J=0}$.} (a) Tracing a contour of the phase boundary between the insulating ($Z<10^{-2}$) and conducting ($Z>10^{-2}$) phase for the wide and narrow bands shows that an OSMT occurs as the bandwidth ratio decreases. The red stars correspond to the same phase boundary obtained with ED for a few bandwidth ratios.(b) Median impurity solver wall-time speedup offered by MOR+EC for the select bandwidth ratios where ED calculations were performed, on a log scale. The impurity solve time is divided into ground state solve (red, GS) and GF evaluation (blue, GF) with the medians for each reported above the bars. Whiskers indicate the $5^\mathrm{th}$ and $95^\mathrm{th}$ percentiles.}
    \label{fig: 2B phase diagram}
    
\vspace{-6pt}
\end{figure*}

We then deploy the ROM to run a much finer scan over $U/W_1$ and $t_2/t_1$ to trace the phase boundary directly (\cref{fig: 2B phase diagram}(a)), rather than inferring it from a handful of ED points. Prior DMFT studies~\cite{Ferrero20052Bdmft,Inaba20052Bdmft, deMedici2005osmt} place the OSMT onset around $t_2/t_1\lesssim0.2$, but our $Z_2\to0$ boundary drops steeply near $t_2/t_1\sim0.15$, while also showing orbital-selective behavior setting in earlier than previously reported. We verified this was not an artifact of hysteresis by reversing the sweep direction in both $U/W_1$ and $t_2/t_1$. The hysteresis sweep showed that the diagram was unchanged and MOR+EC remained accurate throughout. Our leading suspicion is that this reflects our smaller bath ($N_B=3$ per orbital versus the $N_B\geq4$ per orbital typical in prior work), though our specific implementation of the bath-parameter optimization procedure could also shift which DMFT fixed point is reached. In either case, it is the fine resolution this ROM makes affordable that let us see the discrepancy at all.

\cref{fig: 2B phase diagram}(b) shows that the cost of that resolution was comparatively small to that of ED. MOR+EC delivers a median speedup of $42$--$57\times$ for the ground state, $3.3$--$3.5\times$ for the GF, and a total median speedup of $9.5\times$ across the bandwidth ratios where we also ran ED, with wider whiskers on the ground state solves for the same reason worked out in \cref{sec: 1B results}. ED slows down near the transition boundary while MOR+EC's fixed-dimension reduced Hamiltonian does not track that slowdown. The high level of detail in this phase diagram, together with the ease of performing hysteresis to probe phase boundaries, makes the case for extending this approach to $N_B>3$ for each orbital, where full ED becomes considerably more expensive to run at this resolution.

\subsection{Real frequency spectra at no extra cost}

\begin{figure}[t]
    \centering
    \includegraphics[width=0.95\linewidth]{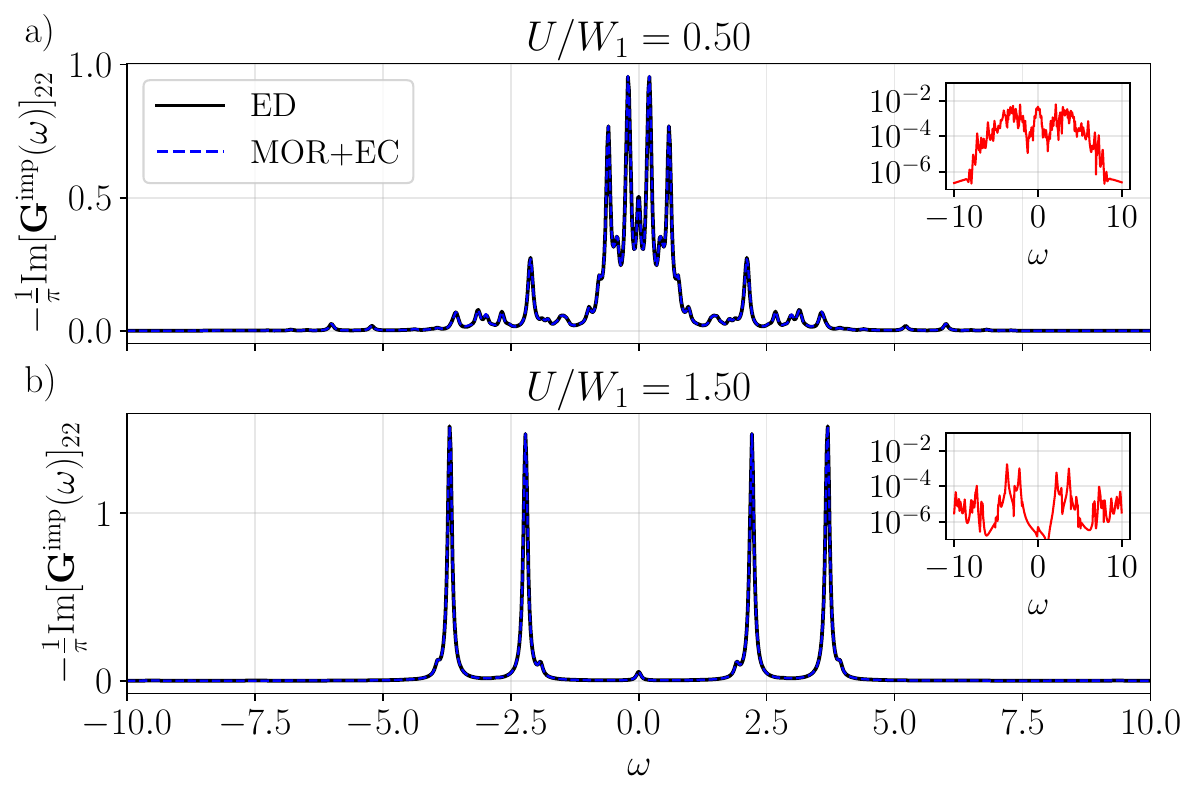}
    \caption{\textbf{Real frequency spectra for the narrow band of the two-band AIM.} The real frequency spectra in the (a) weak and (b) strong interaction regimes are compared to the ED result. The insets show the absolute relative error.}
    \label{fig: real GF comparison}
    
\vspace{-6pt}
\end{figure}

Alongside MOR+EC's ability to efficiently and accurately capture observables through DMFT -- such as the quasiparticle wieght $Z$ and impurity occupation $n_\mathrm{imp}$ -- that are highly sensitive to the accuracy of the GF, the impurity GFs produced have excellent accuracy on both the Matsubara and, remarkably, the real axis. Typically, DMFT is performed on the Matsubara axis where the impurity GF is smooth and amenable to the fitting procedure described in Appendix~\ref{asec: DMFT details}. In our DMFT calculations, we also use the Matsubara axis and purely imaginary interpolation points to build the projections. Because we are interested in the low-energy physics, we heuristically chose a few of the lowest Matsubara frequencies to maximize accuracy near this region, although this choice does not limit the general MOR+EC framework.

While Matsubara GFs are convenient for DMFT stability, real-frequency spectra are typically preferred because they are used to calculate physical observables such as transport coefficients, and the density of states. The task of analytically continuing GFs calculated on the Matsubara axis to the real axis is numerically ill-posed. Methods such as Maximum Entropy~\cite{silver1990maxent,Reymbaut2017maxent,LEVY2017maxent} are typically used, but in our case, this is not necessary.

~\cref{fig: real GF comparison} shows the real frequency spectra for the narrow band. Once the ROM quantities in~\cref{eq: projected model} are computed, evaluating the resolvent in the reduced space on the real axis is straightforward. The MOR+EC solver, while built on the Matsubara axis, works just as well in both the imaginary and real frequency axis. Taking $z$ to be real (with a small, fixed complex broadening, which we take to be $\eta=0.05$), we find excellent agreement on both axes, despite the projections being constructed entirely on the imaginary axis. These spectra are evaluated with the uncompressed ROM, since the compression of Appendix~\ref{asec: compression} only retains what the Matsubara axis requires. The uncompressed ROM is nonetheless competitive on this axis: the Lanczos recursion that ED uses for the GF is typically much longer for real frequencies than for Matsubara frequencies.

Previously, MOR from a data-driven perspective has been used as a vehicle for analytic continuation~\cite{shinaokaCompressingGreensFunction2017,huangBarycentricRationalFunction2025}. However, most of these approaches consider interpolation in the frequency domain only. In our case, MOR+EC also enables parametric interpolation as well. This suggests that MOR-EC could be used as an alternative approach to analytic continuation, as indeed any solver that gives access to the transition 1- and 2-reduced density matrices needed for EC, and to the resolvents needed to compute ROM of the GF, can be used with our methods.

\FloatBarrier

\section{Discussion}\label{sec: discussion}

We have introduced MOR+EC, a framework that constructs reusable reduced-order models for evaluating single-particle GFs by combining eigenvector continuation (EC) in parameter space and rational interpolation using model order reduction (MOR) in frequency space, and applied it as an impurity solver for DMFT. Across two impurity models -- the single-band Anderson model with $N_B=9$ and the two-band model with $N_B=3$ per band -- our reduced-order models constructed with MOR+EC reproduce ED-quality DMFT results at relative errors less than $10^{-3}$ on the impurity GF, while delivering median wall-time speedups of $9.5$--$92\times$. Importantly, in our two-band example, the lower speedup is not due to the multi-band aspect; rather, it results from the total number of orbitals being smaller than in the one-band model ($N=8$ versus $N=10$). Extending this approach to three orbitals with $N_B=3$ bath orbitals per impurity band ($N=12$) is anticipated to produce an even greater speedup. Further, the same projections, despite being constructed entirely on the Matsubara axis, produced real-frequency spectra in agreement with ED. 

From a DMFT standpoint, the defining feature of MOR+EC is that the marginal cost of an additional parameter point decouples from the cost of solving the underlying impurity problem. Once the basis $\mcB$ and the projections $\mbfV,\mbfW$ are built, every subsequent GF solver call reduces to a $k\times k$ generalized eigenproblem and an $m\times m$ resolvent evaluation, with $m\ll d$, the size of the full Hilbert space. Sweeps in chemical potential, bandwidth ratio, or Hund's coupling that take hours of impurity-solver time with ED now take minutes, as demonstrated in~\cref{fig: occ v mu} and~\cref{fig: 2B phase diagram}. Hysteresis loops, which require traversing parameter regions in opposite directions, become inexpensive. Phase boundaries can thus be traced continuously rather than be interpolated between sparse samples, which matters because order parameters extracted from finite grid sweeps are sensitive to the density of samples around the critical point. Studies on sensitivity to perturbations on the model's parameters (Hund's coupling, chemical potential, bath fitting tolerance) also become tractable.

Although our application is DMFT-specific, nothing about the construction of MOR+EC limits it to impurity models: it requires only the ability to compute the snapshots within~\cref{eq: PMOR projections} and that the model's ground state across parameter regimes lies in a low-dimensional subspace, conditions met by a wide range of GF-based problems in condensed matter physics and quantum chemistry. In this, MOR+EC follows a broader trend toward investing once in transferable, high-fidelity surrogates rather than repeatedly performing expensive calculations~\cite{Arsenault2014MLAIM,Aouina2020,Vanzini2022,Giuli2026}, and complements these proposals by targeting the dynamical response directly, rather than the wavefunction or a derived static observable, and by providing a deterministic, error-controllable alternative to data-driven surrogates.

As briefly discussed in~\cref{sec: eigenvector continuation}, the basis $\mcB$ need not be constructed from exact ground states of $\mcH(\param)$. The same generalized eigenvalue problem in~\cref{eq: GEP} can be solved using approximate states drawn from wavefunctions found using Gaussian states or Slater determinants~\cite{hoganEfficientQuantumImplementation2026}, non-orthogonal configuration interaction~\cite{Sundstrom2014,Oosterbaan2018}, coupled cluster~\cite{Shee2019,Zhu2019a}, DMRG~\cite{Nunez2018,Rath2025,Atalar2024}, or variational Monte Carlo~\cite{rigo2025} in the spirit of EC for approximate ground states~\cite{agrawal2025EC}. This opens MOR+EC to using simpler, auxiliary systems: the only requirement being that the chosen solver can provide snapshots that assemble the MOR projections $\mbfV, \mbfW$. The features described here -- once-and-for-all projections, dense phase space exploration, obtaining spectra on the real and imaginary axis -- would then transfer to more complex settings of genuine interest.

While the results presented here are quite promising, there are a few limitations to MOR+EC that may temper our results. First, the speedups reported here are amortized: MOR+EC pays off when the reduced-order model is used across many parameter and frequency evaluations, and for a single self-consistency loop or a small handful of parameters, a direct solve is preferable. Second, the parameter-independent resolvent of Appendix~\ref{asec: param indep resolvent} expands about a fixed reference $\mcH_0$ and retains only the leading term in $\mcH(\param)=\mcH_0+\Dh$. This is accurate whenever $\Dh$ remains small over the parameter range covered by a single reference, as was the case for the models and parameter windows explored here. For systems in which $\mcH(\param)$ departs strongly from any single $\mcH_0$, the expansion would need several local references, at the cost of additional factorizations to build the snapshot vectors. Third, the construction of $\mcB$ typically scales with the number of dimensions of $\param$, exacerbated through the large training grid used to construct the basis; while sparse grid techniques mitigate this~\cite{joe2008constructing, Bungartz_Griebel_2004_sparsegrids}, applications with many simultaneously varied parameters will require more careful strategies. Finally, as described in Appendix~\ref{asec: adpative EC}, we use a Hamiltonian-variance-based cost function to construct the basis $\mcB$, which depends on the 3- and 4-RDMs of the basis vectors. Computing these becomes a dominant cost as $k$ and $Q$ grow.

Several extensions naturally follow from the framework presented here. The most immediate is \textit{locally adapted projections}: rather than a single pair $\mbfV, \mbfW$ for each impurity orbital tailored to the entire parameter domain, a collection of regional projections with overlapping coverage~\cite{benner2016pmorsurvey,Giuli2026} would shrink the effective rank $m$ of each reduced-order model, and the same logic applies to the EC basis as well. A second direction is replacing the heuristic choice of low-energy points restricted to the Matsubara axis as interpolation frequencies $\mbf{z}_r$ with an error-driven selection scheme: \textit{a posteriori} error-estimators for MOR~\cite{feng2024PosterioriErrorEstimation, benner2016pmorsurvey,gugercin2008H2modelreduction,FLAGG2012688} provide convergence guarantees and could reduce the number of interpolation frequencies $r$ needed for a target accuracy. A third direction, alluded to above, is reducing the bottleneck of the variance-based cost function by approximating the 3- and 4-RDMs~\cite{mau2017electroncorrelation,Colmenero1994selfconsistent,Mazziotti2006anti-hermitiancontracted}, which would make the variance criterion cheap at large $k$.

Finally, integrating MOR+EC with approximate solvers such as DMRG, coupled-cluster, or potentially quantum-computing-based methods~\cite{hoganEfficientQuantumImplementation2026, Mejuto-Zaera2023quantumEC, francis2022subspacediagonalizationquantumcomputers, agrawal2025EC} would push the framework closer toward the multi-orbital, materials-realistic regimes that motivate our DMFT demonstration in the first place. Many of these solvers, however, do not expose the impurity Hamiltonian in a form our projection-based construction can act on directly; pairing MOR+EC with them may instead call for data-driven approaches to MOR, which forgo projection altogether and build the reduced-order model purely from measurements of the GF~\cite{ionitaDataDrivenParametrizedModel2014,aretzNestedOperatorInference2025,balickiMultivariateRationalApproximation2025,peherstorferDatadrivenOperatorInference2016}. We view the present work as proof of principle: the impurity models here are tractable by ED but the methodology was designed with larger applications in mind.

\section*{Author Contributions}

NH, AFK, and CMZ developed the framework to combine the ideas of eigenvector continuation and model order reduction. NH, AFK, and CMZ laid the theoretical foundation to combine the MOR+EC solver in the DMFT prescription. CMZ proposed the problem to study with our DMFT calculations (single-band impurity occupation and two-band orbital-selective Mott transition phase diagrams). NH developed the MOR+EC and ED DMFT code and performed all calculations. NH prepared the manuscript and all authors reviewed and provided feedback.

\vfill

\begin{acknowledgments}
We thank Roel Van Beeumen for helpful discussions on model order reduction. The authors also thank Mauricio Rodríguez-Mayorga for their input on approximating 3- and 4-RDMs. We acknowledge the computing resources provided by North Carolina State University High Performance Computing Services Core Facility (RRID:SCR\_022168). We acknowledge the use of Claude Opus 5 for parts of the code development, specifically for debugging, optimization, and implementation of parallelization.
AFK was supported by the U.S. Department of Energy, Office of Science, Office of Advanced Scientific Computing Research under Award Number DE-SC0025430.
\end{acknowledgments}

\section*{Data Availability}
The data and code to reproduce all figures in this study is available at~\cite{morec_code}.

\bibliography{norman,carlos}

\clearpage
\onecolumngrid
\appendix

\renewcommand\thefigure{S\arabic{figure}}  
\renewcommand\thetable{S\arabic{table}}  
\setcounter{figure}{0}

\section{Adaptive construction of the EC basis}\label{asec: adpative EC}

In situations where a sweep through parameters is used, such as found in a convergence procedure used in our DMFT application, we can further leverage the structure of that problem to adaptively update the basis precisely with the vectors that will add information where it is needed. We construct a reduced subspace $\mcB$ that achieves target accuracy across parameter regimes of interest, employing a two-phase adaptive selection procedure consisting of an \emph{offline} global initialization and an \emph{online} trajectory-guided enrichment.

\subsubsection*{Offline Global Initialization}
During the offline phase, we sample the parameter manifold using a quasi-random scrambled Sobol' sequence~\cite{joe2008constructing} to form a training grid $\Xi_\text{global} = \{\param_j\}_{j=1}^M$ ($M \gg 1$) which has global coverage in a specified set of parameter bounds. To minimize expensive full-space exact diagonalizations (ED), we greedily append new basis vectors by searching for parameter points that maximize a surrogate error metric -- the Hamiltonian variance calculated entirely within the reduced space~\cite{herbst2022surrogatemodels}. 

At iteration $n$ (where $n$ is the current rank of $\mcB$), the next parameter point $\param_{n+1}$ is selected according to
\begin{align}\label{eq: hamiltonian variance}
    \param_{n+1} = \underset{\param \in \Xi_\text{global}}{\text{arg\,max}} \ \text{Var}_n(\param)^2,\nonumber
\end{align}
where
\begin{align}
    \text{Var}_n(\param)^2 = &{\varphi^{(0)}_n(\param)}^\dagger \left( \sum_{q, q'}^{Q} p_q p_{q'}\mbf{h}_{qq'} \right) \varphi^{(0)}_n(\param)\nonumber \\ &- \lambda^{(0)}_n(\param)^2.
\end{align}
Here, $\lambda^{(0)}_n(\param)$ and $\varphi^{(0)}_n(\param)$ are the ground state eigenvalue and eigenvector evaluated in the $n$-dimensional reduced basis, and $\mbf{h}_{qq'} = \mcB_n^\dagger \mbf{H}_q \mbf{H}_{q'} \mcB_n \in \reals^{n \times n}$ pre-computes operators incorporating 3- and 4-body density matrices. Because each evaluation in~\cref{eq: hamiltonian variance} is confined to the small reduced basis, evaluating the variance over the training grid $\Xi_\text{global}$ is computationally inexpensive and trivially parallelizable.

Once a point $\param_{n+1}$ is selected, we use full-space ED to yield the exact ground state $|\Phi(\param_{n+1})\rangle$, but an approximate solver may be used to obtain this state provided that it may be represented as a state vector. To prevent numerical ill-conditioning caused by near-linear dependencies, an incremental singular value decomposition (SVD) is performed on $\mcB$ to discard redundant directions before appending the new vector~\cite{Mejuto-Zaera2023quantumEC}. We terminate the subspace growth by setting a threshold $\varepsilon_\off$ on the variance

\begin{align}\label{eq: terminate EC}
    \max (\mathrm{Var}_n(\param)^2)\leq \varepsilon_\off\quad\mathrm{for\ all\ }\param\in\Xi_\mathrm{global}.
\end{align}

\subsubsection*{Online Trajectory Enrichment}
\begin{figure}[t]
    \centering
    \includegraphics[width=0.5\linewidth]{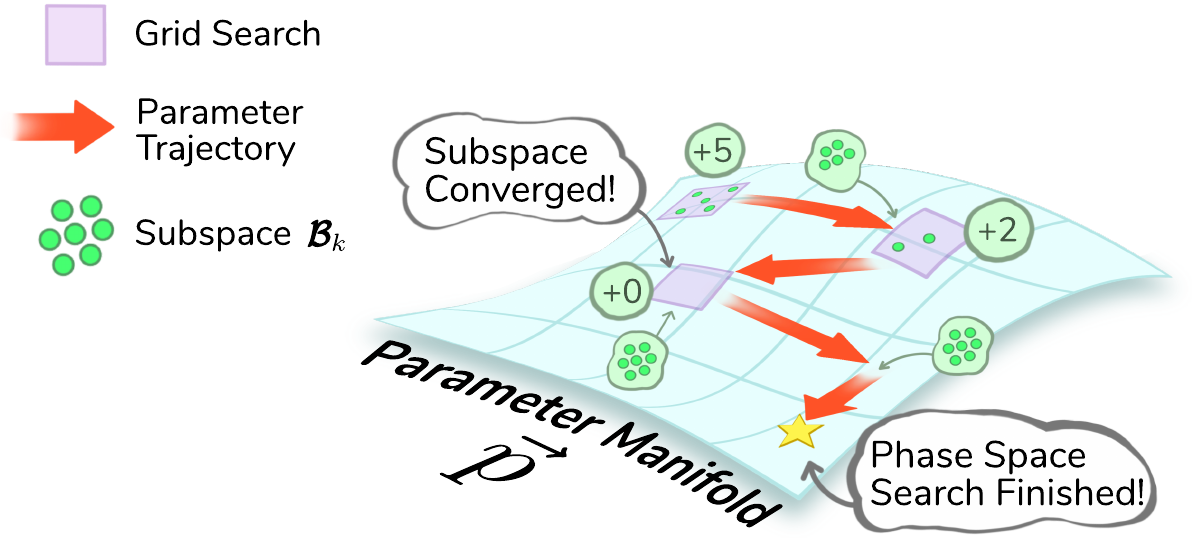}
    \caption{\textbf{Illustration of the online adaptive basis construction for the subspace $\mcB$.} A coarse path in parameter space is traversed (red arrows), with the reduced-basis residual monitored at each step. When the residual exceeds tolerance, a local grid search (purple square) is performed over a local Sobol' sequence. States greedily selected via Hamiltonian variance minimization are calculated in the full space (green circles) and appended to $\mcB$. Once the basis converges along the trajectory, fine-grained evaluations across the path are performed entirely within the reduced basis using \cref{eq: GEP}.}
    \label{fig: adaptive search}
    
\vspace{-6pt}
\end{figure}

When following a specific path through parameter space (red arrows in~\cref{fig: adaptive search}), the global offline basis may not be sufficiently dense along localized regions of the trajectory. During online propagation, we monitor the normalized residual evaluated in the reduced basis,
\begin{align}\label{eq: residual}
    \mathrm{Res}(\param) = \frac{\big\|\, \mcH(\param)\,\mcB\varphi^{(0)}(\param) - \lambda^{(0)}(\param)\,\mcB\varphi^{(0)}(\param)\,\big\|}{\|\mcB\varphi^{(0)}(\param)\|},
\end{align}
where $\|\cdot\|$ denotes the 2-norm.

If $\mathrm{Res}(\param)$ exceeds a specified threshold $\varepsilon_\mathrm{trig}$, an online enrichment event is triggered:
\begin{enumerate}
    \item A local Sobol' grid $\Xi_\mathrm{local}=\{\param_j\}_{j=1}^{M_\mathrm{local}}$ of $M_\mathrm{local} \ll M$ parameter points is generated in a small neighborhood around $\param$ (purple squares in~\cref{fig: adaptive search}).
    \item Local greedy selection is performed over this neighborhood using the reduced-basis variance metric in~\cref{eq: hamiltonian variance} until~\cref{eq: terminate EC} is satisfied for a threshold $\varepsilon_\on$ over the local training grid $\Xi_\mathrm{local}$.
    \item Full-space ED is executed at the chosen local points to obtain new exact vectors (green circles in~\cref{fig: adaptive search}), which are then orthogonalized into $\mcB$ via incremental SVD truncation.
\end{enumerate}

After a single coarse traversal along the parameter path, the enriched basis $\mcB$ fully spans the relevant low-energy subspace. Subsequent fine-grained scans along the path can then be conducted entirely within the reduced basis without further full-space evaluations. For our purposes, we prefer to track the low-energy subspace along a narrow trajectory through parameter space (the DMFT self-consistent trajectory) rather than across the global parameter bounds, so we set $\varepsilon_\off>\varepsilon_\on$. This minimizes the number of vectors added to the basis during the offline global initialization in favor of online enrichment. However, if global accuracy is desired, setting a very low $\varepsilon_\off$ and forgoing online enrichment may suffice.

\section{DMFT self-consistency details}\label{asec: DMFT details}

In the main text, we apply our methods in the Dynamical Mean Field Theory (DMFT) framework~\cite{Kotliar1996,Kotliar2006,Zgid2011,Paul2019} to demonstrate the robustness of MOR+EC for phase space exploration and its stability to parametric changes of the underlying impurity model studied. For the single- and two-band impurity models utilized in this work (see~\cref{eq: 1B local H} and~\cref{eq: 2B local H} for their local Hamiltonian structure), MOR+EC functions as the solver which produces both the ground state and single particle GF -- necessary inputs to the self-consistency equations of DMFT. While these are the expensive ingredients for the DMFT prescription, in this section we detail the surrounding equations to make this text self-contained. In the following, we assume this procedure is performed on the Matsubara axis, $i\omega_n$, with $\omega_n=\frac{(2n+1)\pi}{\beta}$ for fermions and $\beta$ a fictitious inverse temperature. The bath parameters are taken to be real throughout.

On the level of the impurity GF, one extracts the impurity self-energy $\mbf{\Sigma}^\text{imp}$ through Dyson's equation

\begin{align}\label{eq: dyson}
    \mathbf{\Sigma}^\mathrm{imp}(i\omega_n)=(\mbf{G}^{0}(i\omega_n))^{-1}-(\mbf{G}^\mathrm{imp}(i\omega_n))^{-1}
\end{align}

\noindent
where $\mbf{G}^0(i\omega_n)=[i\omega_n\mbfI-\bm{\nu}-\mbf{\Delta}(i\omega_n)]^{-1}$ is the non-interacting GF of the impurity model, hybridized with the bath through

\begin{align}\label{eq: bath hyb}
    [\mbf{\Delta}(i\omega_n)]_{ii} = \sum_b^{N_B} \frac{(V^i_b)^2}{i\omega_n-\epsilon^i_b}.
\end{align}

\noindent
and $\bm{\nu}=\text{diag}(\nu_1,\dots,\nu_{N_I})$ is the one-body matrix of the local part of the Hamiltonian. The chemical potential for the models discussed in this work enters through $\nu_i=\epsilon_i-\mu$.

The self energy of the impurity model is used as a local approximation to the self-energy of the fully correlated model; that is, the lattice self-energy is taken to be momentum independent and is then identified with that of the auxiliary impurity problem,

\begin{align}\label{eq: self consistent}
    \mathbf{\Sigma}^\mathrm{latt}(\mbf{k},i\omega_n)\approx\mathbf{\Sigma}^\mathrm{loc}(i\omega_n)=\mathbf{\Sigma}^\mathrm{imp}(i\omega_n).
\end{align}

\noindent
Using $\mbf{\Sigma}^\text{imp}$, the local GF of the correlated lattice model ($\mbf{G}^\text{loc}$) is evaluated with

\begin{align}\label{eq: local GF}
    \mbf{G}^\text{loc}(i\omega_n)=\int^D_{-D} d\epsilon\;\rho(\epsilon)\left[\mbfI(i\omega_n-\epsilon)-\bm{\nu}-\mbf{\Sigma}^\text{imp}(i\omega_n)\right]^{-1}
\end{align}

\noindent
with the density of states of the Bethe lattice defined as $\rho(\epsilon)=\frac{2}{\pi D}\sqrt{1-(\epsilon/D)^2}$ and $D=2t$ the half-bandwidth. This produces new parameters for the bath ($V^i_b$, $\epsilon^i_b$) through fitting of the hybridization function over the Matsubara grid with the criteria

\begin{align}\label{eq: bath fitting}
    \argmin_{(V^i_b,\text{ }\epsilon^i_b)} \sum_{n=0}^{N_\omega} w_n\left\|\mbf{\Delta}(i\omega_n) - (\mbfI i\omega_n-\bm{\nu}-\mbf{\Sigma}^{\text{imp}}(i\omega_n)-[\mathbf{G}^\text{loc}(i\omega_n)]^{-1})\right\|^2,
\end{align}

\noindent
where $N_\omega$ is the number of frequencies retained and $w_n$ the associated weights. For our implementation, we use a curve fitting function provided through \texttt{scipy.optimize}.

Since the models under consideration do not present any hoppings between impurity orbitals, the bath is parametrized such that each bath orbital couples to a single impurity orbital (cf.~\cref{eq: bath hyb}), and the interaction is of density--density form, the matrix-valued quantities $\mbf{G}^\text{imp}$, $\mbf{\Delta}$, $\mbf{\Sigma}^{\text{imp}}$, and $\mbf{G}^\text{loc}$ are all diagonal in the orbital index. After some iterations, if the self-energy -- and hence the fitted bath parametrization -- does not change, the auxiliary impurity model is self-consistent with the fully correlated lattice model and thus the DMFT is converged.

\section{Numerical details}\label{asec: numerical details}

\subsubsection{Selection of frequency interpolation points}

The choice of interpolation points $\mbf{z}_r$ is crucial. While formal error estimates~\cite{feng2024PosterioriErrorEstimation, benner2016pmorsurvey,gugercin2008H2modelreduction,FLAGG2012688} can guide this choice, for the DMFT application of~\cref{sec: DMFT} we instead select a small number of points heuristically in the low-frequency range, so as to resolve the low-energy properties central to the many-body models considered. These points need not lie on the real axis: all results reported here use points on the Matsubara axis ($\text{Re}[z]=0$). Because the GF is analytic throughout the complex plane away from its poles, a reduced-order model interpolated on the imaginary axis is nonetheless constrained on the real axis, and we find in the main text that it reproduces real-frequency spectral information accurately even at small complex broadening $\eta$.

\subsubsection{Conditioning of projections}

The projections in~\cref{eq: PMOR projections}, and hence the projected quantities $\widetilde\mcH$ and $\widetilde\mbfI$, are of order at most $m=kr$. Regularizing the projections by QR factorization keeps the reduced-order model rank at $m\le kr\ll d$, and during online updates we use incremental QR factorizations to absorb new states into $\mcB$ and their snapshots into $\mbfV,\mbfW$, so that the model remains well conditioned as the basis grows.

\subsubsection{Computational details}

All calculations were implemented in Python. In the main text we benchmark MOR+EC within the DMFT workflow against exact diagonalization (ED); the ED reference data were obtained by diagonalizing the impurity Hamiltonian and evaluating impurity GFs with a Lanczos solver. ED and MOR+EC evaluate the GF with the same continued-fraction Lanczos routine, and the per-call speedups of~\cref{tab:gf_error_channels_timing} and~\cref{sec: 1B results} were measured by giving identical inputs to both solvers in turn on the same compute node. For EC we used dense diagonalization of the reduced matrix $\mbf{h}(\param)$ of order $k$. For the MOR solver, all projections used here are Galerkin ($\mbfV=\mbfW$), so that $\widetilde\mcH(\param)$ is Hermitian and the reduced overlap $\widetilde\mbfI$ is the $m\times m$ identity up to the tolerance of the orthonormalization of $\mbfV$. We therefore work with an orthonormalized $\widetilde\mcH(\param)$, which we monitor and re-orthonormalize as the basis grows, and obtain the reduced-order model GF by Lanczos recursion. This avoids the $\mathcal{O}(m^3)$ dense diagonalization of the order-$m$ reduced-order model at every call to the solver.

The reduced subspace $\mcB$ was built by a combined offline--online procedure. Offline, candidate parameter points were drawn from a global Sobol' grid of $M=1024$ points, and states were appended greedily to $\mcB$ until the maximum variance of~\cref{eq: hamiltonian variance} fell below the tolerance $\varepsilon_\off=10^{-2}$, unless stated otherwise. Online, local enrichment was triggered whenever the normalized residual of~\cref{eq: residual} exceeded a threshold $\varepsilon_\mathrm{trig}$, which we set equal to $\varepsilon_\off$. Each enrichment event built a local Sobol' grid of $M_\mathrm{loc}=16$ points by perturbing the current bath parameters, over which the greedy selection of~\cref{eq: hamiltonian variance} was re-run; states exceeding the online threshold $\varepsilon_\on$, which we vary, were appended to $\mcB$ and the reduced-order model updated before continuing along the parameter trajectory.

Impurity GFs were evaluated at $500$ Matsubara frequencies with $\beta=232$. For MOR+EC these are the frequencies at which the reduced-order model is evaluated, and are distinct from the interpolation points $\mbf{z}_r$ used to construct it. The DMFT loop was terminated once the self-energy changed by less than $10^{-4}$ between iterations, with a maximum of $150$ iterations.

\section{Details on the parameter-independent resolvent}\label{asec: param indep resolvent}

Standard PMOR procedures typically evaluate the snpashots $(z_i \mbfI - \mcH(\param_j))^{-1}$ at the same $\param_j$ as the corresponding subspace vector. Because both the shift $z_i$ and the Hamiltonian $\mcH(\param_j)$ vary, if the interpolation directions are supplied by a subspace $\mcB$ as we have done in this work, this requires factoring $kr$ unique resolvents. This is often the primary bottleneck for large basis sizes, i.e. the parametric dependence on the number of directions grows large. However, using only a single reference Hamiltonian $\mcH_0 = \mcH(\param^{\,0})$, only the $r$ frequency shifts remain, and each of the $r$ factorizations are reused across the $k$ directions in $\mcB$.

This approach is motivated by the observation that PMOR seeks to describe the response of the states $|\Psi(\param)\rangle \approx \mcB\varphi^{(0)}$ to the resolvent, rather than the resolvent in isolation. By expanding the Hamiltonian as $\mcH(\param) = \mcH_0 + \Dh$ and factoring the resolvent $\mbf{R}(\param, z) = (z\mbfI - \mcH(\param))^{-1}$ -- which appears in $\mbfV$, but all derivations also apply to the left hand counterpart $\mbfW$ -- we obtain

\begin{align}\label{eq: expanded resolvent}
    \mbf{R}(\param, z) &= (\mbfI z - \mcH(\param))^{-1}=(\mbfI z-\mcH_0 - \Dh)^{-1}.
\end{align}

\noindent
We define a reference resolvent as $\mbf{R}_0(z) = (z\mbfI - \mcH_0)^{-1}$ and factor it from~\cref{eq: expanded resolvent} to give 

\begin{align}\label{eq: factored R0}
    \mbfI z-\mcH_0 - \Dh =\mbf{R}_0(z)^{-1}-\Dh=\mbf{R}_0(z)^{-1}(\mbfI-\mbf{R}_0(z)\Dh).
\end{align}

\noindent
Inverting~\cref{eq: factored R0} gives
\begin{align}\label{eq: inverted factored R0}
    \mbf{R}(\param, z) &= (\mbf{R}_0(z)^{-1}(\mbfI-\mbf{R}_0(z)\Dh))^{-1}=(\mbfI -\mbf{R}_0(z)\Dh)^{-1}\mbf{R}_0(z),
\end{align}
\noindent
which is well defined when $z$ is not an eigenvalue of either $\mcH(\param)$ or $\Dh$.

Using the geometric series $(\mbfI-\mbf{X})^{-1}=\sum_{n=0}^\infty\mbf{X}^n$ and defining $\mbf{X}=\mbf{R}_0(z)\Dh$,

\begin{align}\label{eq: series expanded resolvent}
    \mbf{R}(\param, z)&=\sum_{n=0}^\infty(\mbf{R}_0(z)\Dh)^n\mbf{R}_0(z)\nonumber\\
    &=\mbf{R}_0(z)+(\mbf{R}_0(z)\Dh)\mbf{R}_0(z)+(\mbf{R}_0(z)\Dh)^2\mbf{R}_0(z)+\dots.
\end{align}

\noindent
Taking $\mbf{b}_j$ to be the $j$-th column of $\mbfB\mcB$, we act this expanded resolvent on $\mbf{b}_j$ to give

\begin{align}\label{eq: series resolvent on bj}
    \mbf{R}(\param, z)\mbf{b}_j=\mbf{R}_0(z)\mbf{b}_j+(\mbf{R}_0(z)\Dh)\mbf{R}_0(z)\mbf{b}_j+(\mbf{R}_0(z)\Dh)^2\mbf{R}_0(z)\mbf{b}_j+\dots,
\end{align}

\noindent
where $\mbf{R}_0(z)\mbf{b}_j$ is exactly the zeroth-order contribution to the snapshot set $\{\mbf{R}(\param, z)\mbf{b}_j\}_{j=1}^k$. Every higher-order term has the form $\mbf{R}_0(z)\mbf{y}$ with $\mbf{y}=(\Dh\mbf{R}_0(z))^n\mbf{b}_j$. The higher-order terms never leave the space $\mbf{R}_0\cdot\mathrm{span}\{(\Dh\mbf{R}_0(z))^n\mbfB\mcB\}$ -- a rational-Krylov-type space seeded by $\mcB$. For many cases where $\param\neq\param^{\,0}$, the form of $\Dh$ is sparse or completely diagonal, as is the case for the impurity models discussed in~\cref{sec: DMFT}, so each higher term mixes only a few new directions which are likely in the span of a sufficiently large $\mcB$ with support over many $\param$.

It's important to note where this approximation struggles the most. The choice to place interpolation frequencies on the Matsubara axis works in our favor for large $\omega_n$: the eigenvalues of $\mcH_0$ and $\mcH(\param)$ sit on the real axis, so each successive term in~\cref{eq: series resolvent on bj} carries a relative factor of order $\sim \|\Dh\|_2/|\omega_n|$, and dropping them is sensible when $z$ is far from the real axis. At low Matsubara frequencies the error of this approximation is concentrated -- exactly where the relevant physics exists. This bounds the neglected terms in the snapshot \emph{directions}, however, not the error of the reduced-order model: \cref{eq: series resolvent on bj} motivates the choice of directions, but once $\mbfV$ and $\mbfW$ are fixed, the accuracy is set by how well $\operatorname{range}(\mbfV)$ collectively spans the $\mbf{R}(\param,z)\mbf{b}_j$ (and likewise for $\mbfW$), not by where the series was truncated. We choose $\mcH_0 = \mcH(\param^{\,0})$ near the center of the parameter manifold explored, minimising the largest deviation $\|\Dh\|_2$ encountered over the domain, and find the error low enough to extract accurate information. In future studies it may be necessary to partition the domain across $n$ reference Hamiltonians, requiring $nr$ factorizations rather than $r$ but still well below $kr$ for $n \ll k$.

\section{Compression of the reduced-order model}\label{asec: compression}

The projections of~\cref{eq: PMOR projections} hold a snapshot for every state of $\mcB$ at every interpolation frequency, so that $m\sim kr$. Along a given scan, however, the ROM is only driven by the vectors $\widetilde{\mbf{b}}(\param)$ built from the reduced ground states $\varphi^{(0)}(\param)$ that the scan visits, and the solutions it generates occupy a subspace of much lower dimension. We remove the unused directions by a proper orthogonal decomposition (POD)~\cite{benner2016pmorsurvey,bonilla2022training,duget2024EC} of the ROM's own solution vectors, with snapshots taken in the frequency domain~\cite{willcox2002balancedpod}.

For each impurity orbital and for each of the particle and hole contributions, we evaluate the reduced solution vectors
\begin{align}\label{eq: reduced snapshots}
    \mbf{x}(\param_j,z_n)=\left(z_n\widetilde\mbfI\pm\widetilde\mcH(\param_j)\right)^{-1}\widetilde{\mbf{b}}(\param_j)\in\complex^{m}
\end{align}
at the final parameters $\param_j$ of each of the $N_\mathrm{c}$ DMFT-converged points of the coarse scan used for online enrichment, and at the $N_z$ lowest Matsubara frequencies $z_n=i\omega_n$. Unlike the interpolation set $\mbf{z}_r=\{z_i\}$ used to generate the snapshots in $\mbfV$ and $\mbfW$, these $N_z$ Matsubara frequencies are chosen densely near the low-frequency region for high resolution of the dynamical information, and the evaluation of~\cref{eq: reduced snapshots} is cheap. $\mbf{x}(\param_j,z_n)$ are the columns of a snapshot matrix $\mbf{X}\in\complex^{m\times N_\mathrm{c}N_z}$. One Hermitian eigendecomposition of the $m\times m$ matrix $\widetilde\mcH(\param_j)$ yields the solutions at all $N_z$ frequencies, so the correlation matrix
\begin{align}\label{eq: correlation matrix}
    \mbf{C}=\mbf{X}\mbf{X}^\dagger=\sum_{j=1}^{N_\mathrm{c}}\sum_{n=1}^{N_z}\mbf{x}(\param_j,z_n)\,\mbf{x}(\param_j,z_n)^\dagger\in\complex^{m\times m}
\end{align}
is accumulated one parameter point at a time, without storing $\mbf{X}$. Its eigendecomposition $\mbf{C}=\mbf{P}\,\mathrm{diag}(\sigma_1^2,\dots,\sigma_m^2)\,\mbf{P}^\dagger$, where $\sigma_1\geq\sigma_2\geq\dots$ are the singular values of $\mbf{X}$, ranks the directions of the reduced space by how strongly the snapshots populate them. We keep the $m'$ eigenvectors with $\sigma_i/\sigma_1>\varepsilon_\mathrm{c}$ as the columns of $\mbf{P}_{m'}\in\complex^{m\times m'}$ and project
\begin{align}\label{eq: compressed model}
    \widetilde\mbfH_q\to\mbf{P}_{m'}^\dagger\,\widetilde\mbfH_q\,\mbf{P}_{m'},\qquad \widetilde{\mbf{b}}(\param)\to\mbf{P}_{m'}^\dagger\,\widetilde{\mbf{b}}(\param),
\end{align}
and likewise for $\widetilde{\mbf{a}}(\param)$, which is equivalent to replacing $\mbfV$ by $\mbfV\mbf{P}_{m'}$ in~\cref{eq: projected model}. The result is again a ROM, now of order $m'$, and is evaluated exactly as before.

The compression acts on the reduced matrices only. It requires $N_\mathrm{c}$ one-time eigendecompositions of order $m$ for each contribution and no operation in the full space, so its cost does not depend on $d$. It leaves the EC basis $\mcB$ and the matrices $\mbf{h}_q$ untouched, so that ground state energies and static expectation values at a given $\param$ are unchanged. For our results, $N_c=72$ and $N_c=55$ for the single- and two-band impurity models' coarse scan. For both cases, we used the first $N_z=40$ Matsubara frequencies $\omega_n=\frac{\pi(2n+1)}{\beta}$ with $\beta=232$ to build the snapshot matrix $\mbf{X}$. These frequency points are hueristically chosen to cover the first $N_z$ points of the Matsubara grid to capture the low-energy features.

The construction above is for a diagonal element of the GF, for which $\mbf{a}(\param)=\mbf{b}(\param)$ and $\mbfV=\mbfW$; these are the only elements needed for the models studied here. For an off-diagonal element, $\mbf{a}(\param)\neq\mbf{b}(\param)$ and $\mbfV\neq\mbfW$, and the reduced transfer function of~\cref{eq: reduced transfer functions} involves a second family of vectors besides the right vectors $\mbf{x}$ of~\cref{eq: reduced snapshots}, the left vectors
\begin{align}\label{eq: left snapshots}
    \mbf{y}(\param_j,z_n)^\dagger=\widetilde{\mbf{a}}(\param_j)^\dagger\left(z_n\widetilde\mbfI\pm\widetilde\mcH(\param_j)\right)^{-1}.
\end{align}
These are the coordinates of $\mbf{a}^\dagger(z\mbfI\pm\mcH)^{-1}$ in the basis $\mbfW$, as $\mbf{x}$ are those of $(z\mbfI\pm\mcH)^{-1}\mbf{b}$ in the basis $\mbfV$, and a basis that represents one family need not represent the other. The compression then uses two correlation matrices, $\mbf{C}_x=\mbf{X}\mbf{X}^\dagger$ and $\mbf{C}_y=\mbf{Y}\mbf{Y}^\dagger$, whose leading eigenvectors $\mbf{P}_{m'}$ and $\mbf{Q}_{m'}$ replace~\cref{eq: compressed model} by
\begin{align}\label{eq: compressed model petrov}
    \widetilde\mbfH_q&\to\mbf{Q}_{m'}^\dagger\,\widetilde\mbfH_q\,\mbf{P}_{m'},\quad
    \widetilde\mbfI\to\mbf{Q}_{m'}^\dagger\,\widetilde\mbfI\,\mbf{P}_{m'},\nonumber\\
    \widetilde{\mbf{b}}(\param)&\to\mbf{Q}_{m'}^\dagger\,\widetilde{\mbf{b}}(\param),\quad
    \widetilde{\mbf{a}}(\param)^\dagger\to\widetilde{\mbf{a}}(\param)^\dagger\,\mbf{P}_{m'},
\end{align}
that is, $\mbfV\to\mbfV\mbf{P}_{m'}$ and $\mbfW\to\mbfW\mbf{Q}_{m'}$, with $m'$ the larger of the two truncation ranks so that the compressed ROM remains square. Writing $\mbf{x}'$ and $\mbf{y}'$ for the solutions of the compressed ROM expressed in the uncompressed coordinates, the error of its transfer function is
\begin{align}\label{eq: compression error}
    \widetilde\gamma-\widetilde\gamma'=\left(\mbf{y}-\mbf{y}'\right)^\dagger\left(z\widetilde\mbfI\pm\widetilde\mcH\right)\left(\mbf{x}-\mbf{x}'\right),
\end{align}
the product of the errors of the two families, which is why both must be represented. The compressed overlap $\mbf{Q}_{m'}^\dagger\widetilde\mbfI\mbf{P}_{m'}$ is no longer the identity, and its conditioning has to be controlled as for the uncompressed projections (Appendix~\ref{asec: numerical details}).

\end{document}